\documentclass[11pt,a4paper]{article}
\usepackage[utf8]{inputenc}
\usepackage{graphicx,amsmath,amsthm,latexsym,amssymb}
\usepackage{float,epsfig,multirow,rotating,times}
\usepackage{multirow,subfig,pbox}
\usepackage{fixltx2e}
\usepackage{upgreek,wrapfig}
\usepackage{comment}
\usepackage{mathrsfs}
\usepackage{xcolor}
\usepackage{bm}
\usepackage{algorithm}
\usepackage[noend]{algpseudocode}
\usepackage{enumerate}
\usepackage[left=1in, right=1in, top=1in, bottom=1in]{geometry}
\usepackage[colorlinks,citecolor=blue,urlcolor=blue,filecolor=blue]{hyperref}
\usepackage{authblk}
\usepackage{natbib}
\newcommand {\ctn}{\citep}
\newcommand {\ctp}{\citet}

\newtheorem{definition}{Definition}[section]
\newtheorem{theorem}{Theorem}[section]

\newcommand{\abs}[1]{|#1|}

\newcommand{\bd}{\boldsymbol{d}}

\newcommand{\btheta}{\boldsymbol{\theta}}
\newcommand{\bLambda}{\boldsymbol{\Lambda}}

\newcommand{\bTheta}{\boldsymbol{\Theta}}
\newcommand{\bgamma}{\boldsymbol{\gamma}}
\newcommand{\bSigma}{\boldsymbol{\Sigma}}

\newcommand{\bpsi}{\boldsymbol{\psi}}
\newcommand{\bepsilon}{\boldsymbol{\epsilon}}

\newcommand{\bmu}{\boldsymbol{\mu}}

\newcommand{\bvarrho}{\boldsymbol{\varrho}}

\newcommand{\btau}{\boldsymbol{\tau}}
\newcommand{\brho}{\boldsymbol{\rho}}
\newcommand{\bnu}{\boldsymbol{\nu}}
\newcommand{\bsigma}{\boldsymbol{\sigma}}

\newcommand{\bV}{\boldsymbol{V}}

\newcommand{\bM}{\boldsymbol{M}}

\newcommand{\bI}{\boldsymbol{I}}

\newcommand{\bP}{\boldsymbol{P}}

\newcommand{\bR}{\boldsymbol{R}}
\newcommand{\bW}{\boldsymbol{W}}

\newcommand{\bx}{\boldsymbol{x}}
\newcommand{\bX}{\boldsymbol{X}}

\newcommand{\bY}{\boldsymbol{Y}}
\newcommand{\bZ}{\boldsymbol{Z}}

\newcommand{\bzero}{\boldsymbol{0}}
\newcommand{\bone}{\boldsymbol{1}}

\newcommand{\postp}{P_{\bpsi|\bZ}}
\newcommand{\pexp}{E_{\bpsi|\bZ}}

\newcommand{\dec}{\boldsymbol{d}}

\newcommand{\mvn}{\mathcal{MVN}}
\newcommand{\mn}{\mathcal{MN}}
\newcommand{\bgp}{\mathcal{BGP}}

\DeclareMathOperator*{\argmax}{argmax}

\DeclareMathOperator*{\vecc}{vec}
\newcommand{\vecop}[1]{\vecc\left( #1\right)}
\DeclareMathOperator*{\diag}{diag}

\begin{document}
\title{\vspace{-0.8in}
	A Novel Bayesian Multiple Testing Approach to Deregulated miRNA Discovery Harnessing Positional Clustering}
\date{\vspace{-0.5in}}

\author[1]{Noirrit Kiran Chandra\thanks{noirritchandra@gmail.com}}
\author[2]{Richa Singh \thanks{richasngh193@gmail.com}}
\author[1]{Sourabh Bhattacharya\thanks{bhsourabh@gmail.com}}
\affil[1]{Interdisciplinary Statistical Research Unit, Indian Statistical Institute, Kolkata.}
\affil[2]{Human Genetics Unit, Indian Statistical Institute, Kolkata.}

\renewcommand\Authands{ and }
\maketitle
\begin{abstract}
MicroRNAs (miRNAs) are small non-coding RNAs that function as regulators of gene expression. In recent years, there has been a tremendous and growing interest among researchers to investigate the role of miRNAs in normal cellular as well as in disease processes. Thus to investigate the role of miRNAs in oral cancer, we analyse the expression levels of miRNAs to identify miRNAs with statistically significant differential expression in cancer tissues.

In this article, we propose a novel Bayesian hierarchical model of miRNA expression data. Compelling evidences have demonstrated that the transcription process of miRNAs in human genome is a latent process instrumental for the observed expression levels. We take into account positional clustering of the miRNAs in the analysis and model the latent transcription phenomenon nonparametrically by an appropriate Gaussian process.

For the testing purpose we employ a novel Bayesian multiple testing method where we mainly focus on utilizing the dependence structure between the hypotheses for better results, while also ensuring optimality in many respects. Indeed, our non-marginal method yielded results in accordance with the underlying scientific knowledge which are found to be missed by the very popular Benjamini-Hochberg method.
\\[2mm]
\end{abstract}
\maketitle

\section{Introduction}
\label{sec:intro}
MicroRNAs (miRNAs) are small 
non-coding, single-stranded RNAs that function in the post-transcriptional regulation of gene expression. 
There has been a tremendous and growing interest among researchers to investigate the role of miRNA in normal as well as in disease processes over the last decade. 
Aberrant expressions of these tiny regulatory RNA molecules do have direct functional implications in carcinogenic transformation or in further tumour progression towards lethal metastatic forms
\ctn{iorio2012, jansson2012}. From rigorous genetic studies over the years it has been found that miRNAs can function as both pro-tumorigenic 
and anti-tumorigenic molecules \ctn{zhang07, shenouda2009} depending upon the genes which they target. 

Considering the important role miRNAs play in tumorigenic processes, we try to identify and study miRNAs which might play significant roles in head and neck carcinoma. According to \ctp{who2015}, a high percentage of Indian population are regular and direct tobacco users and tobacco users have a significantly higher risk of cancer development. Gingivo buccal squamous cell carcinoma (GBSCC) is one of the most prevalent among tobacco users and categorized as one of the most prominent type of head and neck carcinoma as well. Head and neck squamous cell carcinoma stands as the 
fifth most common malignancy worldwide \ctn{jemal2005} and widely found among the tobacco users of India. Thus we analyze miRNA expression data corresponding to cancer and normal tissues 
collected from GBSCC patients.

We propose a new Bayesian hierarchical model in this paper for differential expressions of the miRNAs harnessing information regarding their positional clustering. Subsequently we apply 
a novel Bayesian multiple testing methodology to detect miRNAs with statistically significant differential expressions. The specific data that we analyse in this context is detailed in 
Section \ref{subsec:data} and in Section \ref{subsec:testing_modelling} we provide an overview of
the Bayesian model and the multiple testing method that we employ in order to detect significant miRNAs. 


\subsection{The data details}
\label{subsec:data}
We used publicly available de-identified 18 patients' 531 miRNA expression data from \ctp{nde14}. They generated these miRNA expression data from cancer tissues
and normal tissues by a high throughput real time polymerase chain reaction (RTPCR) assay (TaqMan Low Density Arrays (TLDA), Applied Biosystems). 
Following the notations of \citet{livak2001} we write, $Ct =$ Cycles at which the PCR product quantity reaches a defined threshold, 
$\Delta Ct =$ Centred  $Ct$ values with respect to the geometric mean of expression of 3 most stable endogenous control miRNAs in that tissue and
$\Delta\Delta Ct = \Delta Ct$\textsubscript{of a miRNA in cancer tissue } $-$ $\Delta Ct$\textsubscript{ of that miRNA in control tissue}. 
The $\Delta Ct$ values were recorded for case and control tissues for the 531 miRNAs across the 18 patients to derive the $\Delta\Delta Ct$ values. By design, some miRNA assays were done in duplicate; after the removal of such duplicates and technical control miRNA data points, 522 unique miRNAs $\Delta\Delta Ct$ values or expression deregulation data remained. It is important to note that, although the $Ct$ values are positive, the $\Delta Ct$ values need not be so as they are centred with respect to endogenous control. In this article, the $\Delta Ct$ value of each miRNA is referred to its respective expression level. 
Now note that for some miRNA $\abs{ \Delta\Delta Ct}>1$ implies that the number of cycles needed to reach the predefined threshold is $\pm1$ in cancer versus normal tissues. Therefore the expression of the particular miRNA is 2 fold lower or higher in cancer tissues compared to its paired normal. If a tumour has at least 2 fold relatively higher expression as compared to its normal counterpart, it is usually referred to as up regulated (-1 or less $\Delta\Delta Ct$ value) and vice versa. Such up and down regulation of a critical miRNA can be interpreted as a biologically significant deviation from its normal quantity and could result in a possible significant functional impact.

\subsection{A brief overview of our Bayesian hierarchical model and the multiple testing problem}
\label{subsec:testing_modelling}
Our interest lies in detecting those miRNAs that significantly influence the disease concerned in the population, based on an appropriate and new Bayesian model, a novel multiple testing
method, and the available sample data set. We develop the model in three stages. We first propose an additive model for the expression level of the miRNAs associated with Gaussian error terms in
Section \ref{sec:model}. 
In Section \ref{sec:matrix_variate}, we model the transcription process of the miRNAs by a latent Gaussian process instrumental for the observed expression levels. Biological admissibility 
of such modelling is supported by a number of studies discussed in the same section. The latent process also realistically accounts for the dependence between the miRNAs that occurs in the course of the transcription process. In Section \ref{sec:generalized_model} we generalize the model to account for miRNAs expressed from different strands of different chromosomes and also formally state the multiple testing problem of interest. 
We also validate and check the goodness-of-fit of the proposed model using leave-one-out cross-validation based on relevant posterior predictive distributions.

In Section \ref{sec:non_marginal} we describe the novel Bayesian multiple testing method proposed by \ctp{chandra2017} to detect significantly deregulated miRNAs. In their multiple testing method, the dependence structure between the miRNA is explicitly exploited. Application of their procedure to this problem yielded insights and 
information that were unrevealed in the previous study of the same data set
by \citet{nde14}; the details are presented in Sections \ref{sec:application} and \ref{sec:comparison}.
Finally, we summarize our contributions and make concluding remarks in Section \ref{sec:conclusion}.


\section{Modelling the data and formulation of the multiple testing problem}	
\label{sec:model}
Let $\bY_j=(Y_{j1},\ldots,Y_{jm})^T$ be the $\Delta Ct$ values of $m (=522)$ miRNAs corresponding to cancer tissues of the $j^{th}$ individual  and 
$\widetilde{\bY}_j=(\widetilde{Y}_{j1},\ldots,\widetilde{Y}_{jm})^T$ 
be the $\Delta Ct$ values from the normal tissues of the same individual, where the suffix $^T$ denotes transpose of the corresponding vector.  
We also assume that $\bY_j$ and $\tilde{\bY}_j$ are both independently and identically distributed for  
$j=1,\ldots, n (=18)$, and that $\bY_j$ and $\tilde{\bY}_j$ are independent of each other. Specifically, we assume that for $j=1,\ldots,n$ and $i=1,\ldots,m$,
\begin{align}
Y_{ji}=\theta_i+\epsilon_{ji}\mbox{ and }
\widetilde{Y}_{ji}=\widetilde{\theta}_i+\widetilde{\epsilon}_{ji},\label{eq: model2}
\end{align}
where $\btheta=(\theta_1,\ldots,\theta_m)^T$ and $\widetilde{\btheta}=(\widetilde{\theta}_1,\ldots,\widetilde{\theta}_m)^T$ are the mean expression levels of the miRNAs 
for case and control respectively and $\bepsilon_j=(\epsilon_{j1},\ldots,\epsilon_{jm})^T$, $\widetilde{\bepsilon}_j=(\widetilde{\epsilon}_{j1},\ldots,\widetilde{\epsilon}_{jm})^T$ are the 
random errors. We assume that
\begin{align}
\bepsilon_j\overset{iid}{\sim} \mathcal{MN}(\bzero, \bLambda_m)\mbox{ and }
\widetilde{\bepsilon}_j\overset{iid}{\sim} \mathcal{MN}(\bzero, \widetilde{\bLambda}_m),
\end{align}
and $\bepsilon_j$, $\widetilde{\bepsilon}_j$ are independent for all $j$. In the above, $\mathcal{MN}(\bzero, \bLambda_m)$ and $\mathcal{MN}(\bzero, \widetilde{\bLambda}_m)$
stand for multivariate ($m$-variate) normal with mean $\bzero$ (the $m$-component vector with all elements zero) and positive definite covariance matrices $\bLambda_m$ and 
$\widetilde{\bLambda}_m$, respectively. 

We consider further dependence structure on $\btheta$ and $\widetilde{\btheta}$ by a bivariate Gaussian process to account their inherent dependence structure. Also note that, in contrast with the usual white noise errors, here we consider errors with dependence structures, which begs some explanation. This is discussed at the end of Section \ref{subsec:matrix_variate_specifics}.

\section{Modelling the transcription process using bivariate Gaussian process}
\label{sec:matrix_variate}
\subsection{Motivation behind the latent process perspective}
\label{subsec:motivation}
There has been quite substantial advancement in statistical modelling of miRNA expression data and identification of differentially expressed miRNA genes. 
Clustering approaches have been extensively used to
find coexpressed genes. Genes in the same positional clusters are anticipated to be coexpressed together. Clustered expression analysis of miRNAs also helps in detecting biomarkers which otherwise would have been 
difficult due to the low intensity of their individual expression levels. Clustering of genes often reduces complexity and also yields greater power. 
Detailed review on popular clustering-based methods can be found in \ctp{wang2015, conesa2016}.


However, the existing methods do not take into account any dependence structure which occurs through the transcription process of the miRNAs. Many of the miRNAs are located in the intronic regions 
of host genes but may have their own promoter regions \ctn{naturereview2017}. Apart from the intronic regions of protein coding genes miRNAs are commonly found to be located in the clusters in 
the different intergenic regions of the genome \ctn{macfarlane2010}. Since miRNAs are found to be in the genomic clusters their transcription might be regulated under the same promoter and 
likely to be coexpressed as reported for a few miRNAs \ctn{promoter2008}. Studies demonstrate that expression levels of different genes are not merely random, but the neighbouring genes 
in the genome have the tendency to be coexpressed together  \ctn{colocalized2008}. These studies provide fair reasons to believe that transcription of the genes in the genome is not just 
a random phenomenon but are subject to an underlying latent process. We believe that modelling the latent process and incorporating it into the analysis is necessary to account for the 
dependence between genes in a more accurate way to obtain reliable and interpretable results. In light of this discussion we propose and describe our approach of modelling the transcription 
process of miRNAs.

\subsection{Specifics of the bivariate Gaussian process modelling of the latent transcription process}
\label{subsec:matrix_variate_specifics}

An important aspect of our modelling strategy, related to considering the transcription phenomenon as a latent stochastic process, is incorporation of positional clusterings of the miRNAs and their genomic coordinates into our model. miRNAs under the same positional cluster are likely to be coexpressed together. Now it is to be noted that there are 22 pairs of autosomes 
(Chr1 through Chr22) and a pair of sex chromosomes (XX/XY) in a human cell and each of them have two strands (we denote the two strands by ``+" and ``-"). Thus there are 46 strands in total. As regards the discussion in the previous section, miRNAs located close to each other on the genome, 
that is, neighbouring miRNAs located on the same strand are expected to be regulated under the same promoter. Therefore, within each strand nearby miRNAs can be regarded 
to be in the same positional cluster. To incorporate this biological information statistically, we model the latent transcription process of miRNAs by a stationary \textit{Gaussian process}. 
In stationary Gaussian process structures, correlation between two miRNA expressions is inversely proportional to their distance provided the miRNAs are expressed from the same strand.

For miRNAs lying in different chromosomes as well as in different strands of the same chromosome,  are not regulated by the same promoter. Therefore, the transcription processes of miRNAs for different strands are considered independent and hence 
we do not put any dependence structure  between the strands {\it a priori}. 

Now, $\btheta$ and $\widetilde{\btheta}$ in (\ref{eq: model2}) are the mean expression levels of the miRNAs corresponding to case and control tissues respectively. 
Since the case and control values for each miRNA are paired observations collected from cancer and normal tissues of the same individual, a biological association 
should be present within each pair. We account for this dependence, as well as the dependence between the miRNAs induced by the transcription process, by assuming 
a \textit{bivariate Gaussian process} $(\bgp)$ associated with $\btheta$ and $\widetilde{\btheta}$ within each strand.
For $i=1,\ldots,k$, where $k=46$ is the total number of strands, consider the $i^{th}$ strand and let 
$\theta^{(i)} (x)$ and $\widetilde{\theta}^{(i)}(x)$ be the expression levels of the miRNA expressed from coordinate $x$ of the $i^{th}$ strand for case and control respectively. 
The genomic coordinates of the miRNAs were obtained from  miRBase \ctn{griffiths2006}. Then
\begin{equation}
\left[\begin{matrix}
\theta^{(i)}(\cdot) \\
\widetilde{\theta}^{(i)}(\cdot)
\end{matrix} \right] \sim\mathcal{BGP}\left(M^{(i)}(\cdot),c^{(i)}(\cdot,\cdot),U^{(i)}\right),
\end{equation}
where, for any $x$, $M^{(i)}(\cdot)=\begin{bmatrix}\mu^{(i)}(\cdot)\\\widetilde{\mu}^{(i)}(\cdot)\end{bmatrix}$ is the bivariate mean function, $U^{(i)}$ is a $2\times 2$ 
positive definite matrix, and $c^{(i)}(\cdot,\cdot)$ is a positive definite stationary covariance function. Since {\it a priori} information on the mean functions $\mu$ and $\widetilde{\mu}$ are unavailable, 
we consider the same mean function for $\theta^{(i)}$ and $\widetilde{\theta}^{(i)}$, that is, we assume $\widetilde{\mu}=\mu$. 
As regards $c^{(i)}(\cdot,\cdot)$, we have considered the Mat\'ern covariance function here with their own set of different hyperparameters for different strands. 
Technical details on $\bgp$ and Mat\'ern covariance function are briefly discussed in \ref{sec:bgp}.

\begin{figure}
	\centering
	\includegraphics[scale=.5]{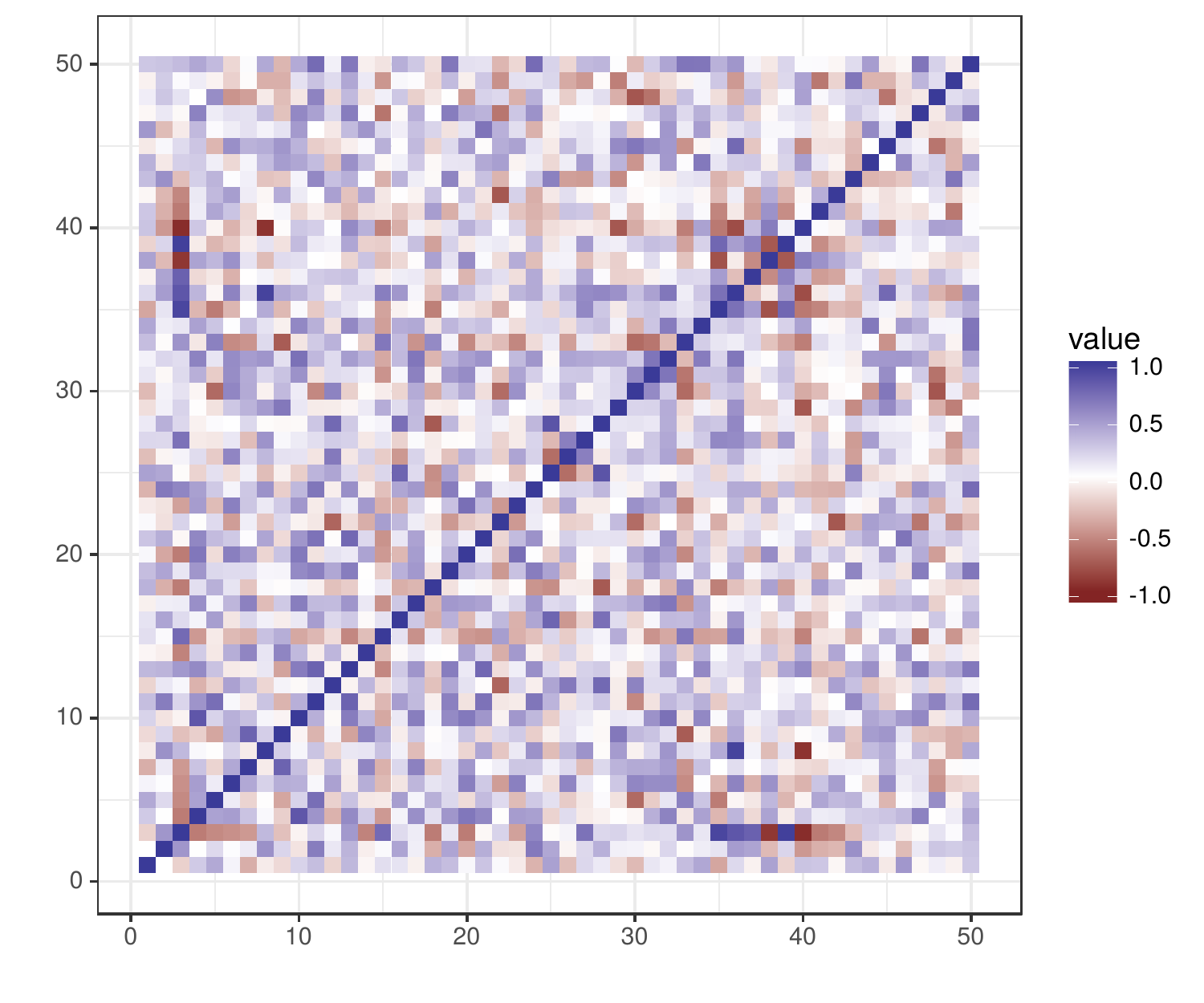}
	\caption{Correlation heatmap of miRNA expressions.}
	\label{fig:heatmap}
\end{figure}

Before we proceed to generalize our model, it is worth shedding light on an important aspect
of our model. Indeed, recall that in Section \ref{sec:model} we 
considered errors with dependence structures, rather than the white noise error traditionally assumed. Note that the mean expression levels of the miRNAs associated with
different strands of the chromosomes are independent of each other as they are controlled independently by different promoters. However, exploratory data analysis exhibited 
significant correlation between the observed expression
levels, even across the strands (see Figure \ref{fig:heatmap}). This suggests that there are hitherto unexplored biological factors responsible for such correlations. The structured error distributions account for
such correlations, making the data dependent, within and across the clusters, ensuring consistency with the exploratory analysis.

\section{Generalization of the model to account for miRNAs expressed from different strands of different chromosomes}
\label{sec:generalized_model}
\begin{scriptsize}
	\begin{align}
	Chr1-&
	\begin{cases}
	miR-135b-5p\\			
	\mathbf{\mathbin{\textcolor{blue}{miR-181a-5p}}}\\
	miR-199a-3p\\
	\mathbf{\mathbin{\textcolor{red}{miR-194-5p}}}\\
	\ldots
	\end{cases} &
	Chr2-&
	\begin{cases}
	\mathbf{\mathbin{\textcolor{red}{miR-194-5p}}}\\
	miR-216b-5p\\
	miR-217\\
	miR-216b-5p\\
	\ldots
	\end{cases}
	ChrX-&
	\begin{cases}
	\mathbf{\mathbin{\textcolor{olive}{let-7f-5p}}}\\
	miR-106a-5p\\
	miR-18b-5p\\
	miR-19b-3p\\
	\ldots
	\end{cases}		
	&		Chr9+&
	\begin{cases}
	let-7f-1-3p\\
	\mathbf{\mathbin{\textcolor{blue}{miR-181a-5p}}}\\
	miR-126-5p\\
	\mathbf{\mathbin{\textcolor{olive}{let-7f-5p}}}\\
	\ldots
	\end{cases}\label{eq:trans2}
	\end{align}	
\end{scriptsize}
Notably, some miRNAs can be expressed from two or more genomic regions also (the coloured miRNAs in (\ref{eq:trans2}) exhibit such instances). For such miRNAs, it is not possible to determine the exact expression level corresponding to each region separately from the results of the 
PCR experiment. As a result we only have total expression levels of the miRNAs in the dataset. For these types of miRNAs, we make some minor modifications to the model.  Let $r_l$ be the number of 
miRNAs in the $l^{th}$ strand and $\bx_l=(x_{l_1},\ldots,x_{l_{r_l}})$ be their corresponding coordinates in the genome. Note that the miRNAs which are transcribed from several genomic 
locations also appear in more than one positional cluster according to their genomic coordinates. Hence, the same miRNA is reported in multiple strands and therefore $\sum_{l=1}^k r_l=L>m$. 

Now consider $miR-181a-5p$ as an illustrating example. Let $Y_{ji}$ be its observed expression level for the $j^{th}$ individual from case tissues. 
From (\ref{eq:trans2}) we see that it is transcribed from two genomic locations. Let $miR-181a-5p$ appear in the $l^{th}$ and the $k^{th}$ strand with
genomic coordinates $x_{l_{r_l}}$ and $x_{k_{r_k}}$, respectively. 
Then we consider the following modification: 
\begin{align}
Y_{ji}=\theta^{(l)}(x_{l_{r_l}})+\theta^{(k)}(x_{k_{r_k}})+\epsilon_{ji}.
\end{align}

In general, suppose a miRNA is in $q$ strands and $x_{l_1},\cdots, x_{l_q}$ be the corresponding genomic coordinates. Then
\begin{align}
Y_{ji}=\theta^{(l_1)}(x_{l_1})+\cdots+ \theta^{(l_r)}(x_{l_q})+\epsilon_{ji}.
\end{align}

For miRNAs which are transcribed from some unique genomic location, we assume the following model:
\begin{align}
Y_{ji}=\theta^{(l)}(x_l)+\epsilon_{ji}.
\end{align}

Let $\bX=(\bx^T_1,\cdots,\bx^T_k)^T$ be the set of all coordinates of the miRNAs across all strands and let $\btheta(\bX)= (\btheta^{(1)}(\bx_1)^T,\cdots,\btheta^{(k)}(\bx_k)^T)^T$ be 
the corresponding mean expression levels, where for any $l=1,\ldots,k$, $\btheta^{(l)}(\bx_l)=(\theta^{(l)}(x_{l_1}),\ldots,\theta^{(l)}(x_{l_{r_l}}))^T$. 
Then
\begin{align}
\bY_j= \bP\btheta(\bX)+\bepsilon_j, 
\end{align}	 
for a known matrix $\bP$ of order $m\times L$.

Similarly for expression levels corresponding to control tissues we have
\begin{align}
\widetilde{\bY}_j= \bP\widetilde{ \btheta} (\bX)+\widetilde{\bepsilon}_j. 
\end{align}

However, we are mainly interested in identifying miRNAs which are significantly deregulated in case tissues as compared to control tissues. Therefore, it is sufficient to work with the 
differential expressions only, instead of dealing with both $\btheta$ and $\widetilde{\btheta}$. We define	
\begin{equation}
\bZ_j=\bY_j-\widetilde{\bY}_j\text{ and } \bpsi=\bP(\btheta(\bX)-\widetilde{\btheta}(\bX)). \label{eq:newmodel}
\end{equation}  
Then $\bZ_j=\bpsi+\btau_j$ where $\btau_j=\bepsilon_j-\widetilde{\bepsilon}_j$. 

Then in terms of $\bpsi$, in accordance with the objective discussed in Section \ref{subsec:data}, the hypothesis testing problem can be framed as
\begin{equation}
H_{0i}:\abs{\psi_i} \leq 1  \hbox{ versus } H_{1i} : \abs{\psi_i}> 1,~i=1,\ldots,m.\label{eq:hypotheses2}
\end{equation}
Working with differential expression values has certain advantages over considering both the case and control values. Although it is not possible to infer about 
$\btheta$ and $\widetilde{\btheta}$ separately from the $\bZ$ values, for the testing problem (\ref{eq:hypotheses2}), considering the $\bZ$ values is sufficient (notably, $\bZ$ values are actually the $\Delta\Delta Ct$ values defined in Section \ref{subsec:data}). 
It not only simplifies the model, but also reduces the number 
of parameters, thus significantly improving the performance of our Markov chain Monte Carlo (MCMC) strategy discussed subsequently.

For the re-framed model (\ref{eq:newmodel}), we need to obtain the distribution of $\btheta-\widetilde{\btheta}$, where $\btheta$ and $\widetilde{\btheta}$ are associated
with the bivariate Gaussian process. The following theorem, the proof of which is straightforward, gives the desired distribution:
\begin{theorem}
	\label{theorem:gp}
	Let $\btheta$ and $\widetilde{\btheta}$ be associated with a bivariate Gaussian process given by
	\begin{equation}
	\left[\begin{matrix}
	\theta(\cdot) \\
	\widetilde{\theta}(\cdot)
	\end{matrix} \right] \sim\bgp\left(\left[\begin{matrix}
	\mu(\cdot)\\
	\mu(\cdot)
	\end{matrix} \right],c(\cdot,\cdot),U \right).
	\end{equation}		
	Define $\gamma(x)=\theta(x)-\widetilde{\theta}(x)$ for all $x$. Then 
	\begin{equation}
	\gamma(\cdot) \sim \mathcal{GP}\left(0,|U|\times c(\cdot,\cdot)\right). \label{eq:gp}
	\end{equation}
	where $|U|$ is the determinant of $U$ and $\mathcal{GP}$ is Gaussian process.
\end{theorem}

Following Theorem \ref{theorem:gp}, we see that given the Mat\'ern hyperparamaters $\bsigma,\bnu$ and $\brho$, $\bpsi$ follows multivariate normal distribution \textit{a priori} (details presented in \ref{sec:psidist}).

Now, recall that
\begin{align}
\bZ_j&=\bpsi+\btau_j\text{ and}\\	
\btau_j&\overset{iid}{\sim} \mn(\bzero, \bSigma),\text{ where } \bSigma=\bLambda+\widetilde{\bLambda},\mbox{ for all }j=1,\ldots, n.
\end{align}

Here $\bSigma,~\bvarrho,~\bnu,~\brho$ all are unknown parameters. It is to be noted that
\begin{equation}
\bZ_j|\bpsi,\bSigma \overset{iid}{\sim} \mn(\bpsi, \bSigma),
\end{equation}
that is, $\bSigma$ is the conditional dispersion matrix of the $\bZ_j$s. We put Inverse-Wishart prior on $\bSigma$. Appropriate prior distributions are also considered over the Mat\'ern hyperparameters. In \ref{subsec:prior_hyperparameters}, we discuss our choice of prior distributions in detail.

Once the Bayesian hierarchical model is complete, samples from the posterior distribution of $\bpsi$ conditional on observed data $\bZ$ requires to be generated to carry out the multiple hypothesis testing problem in (\ref{eq:hypotheses2}). We do this by the fast and efficient Transformation based Markov chain Monte Carlo (TMCMC) method \ctn{Somak14}. We also check the goodness-of-fit of our proposed model by means of the leave-one-out cross-validation by computing appropriate posterior predictive distribution. Details on the TMCMC method and goodness-of-fit results are discussed in \ref{sec:tmcmc}.

\section{A new Bayesian non-marginal multiple testing proposal}
\label{sec:non_marginal}

\subsection{Motivation for non-marginal Bayesian multiple testing procedure}
\label{subsec:motivation_non_marginal}
An analysis of the expression data is done by \cite{nde14}, though the transcription process of the miRNAs was not considered in their model. They performed individual 
$t$-test for all the miRNAs and applied the Benjamini-Hochberg (BH) method \ctn{Benjamini95} for multiplicity correction. The BH method is a $p$-value based procedure of multiplicity adjustment; however, in the BH procedure, the dependence structure between the test statistics is not utilized. Although validity of the BH procedure holds under positive dependence \ctn{by01}, 
for negative dependence the usefulness of BH is unclear. 

\ctp{finner2007, efron2007} discussed the effect of dependence between test statistics, among others. In Bayesian approaches, a natural dependence occurs between hypotheses through hierarchical modelling and multiplicity correction is taken care of to 
some extent \ctn{scott10}. Loss function based approaches have been discussed by \ctp{muller04}. It is interesting that in the aforementioned Bayesian works, positive dependence,
unlike the popular BH procedure, is not required.

However, most of the multiple testing methods, either classical or Bayesian, primarily focus on the validity of the test procedure in the sense of controlling $FDR$ or oracle property corresponding to some loss function, 
whereas exploiting the information provided by the dependence structure might yield efficient closer to truth inference \ctn{sun2009}. When the decisions are not directly (deterministically) 
dependent, information provided by the joint structure inherent in the hypotheses are somewhat neglected by the marginal multiple testing approaches, even though the data 
(and the prior in the Bayesian case) possess dependence structure(s).

Keeping in mind the necessity of exploiting the dependence structure directly in the methodology, a novel Bayesian non-marginal multiple testing procedure is devised by \ctp{chandra2017}. 
The method, which is based on new notions of error and non-error terms, substantially enhances efficiency by judicious exploitation of the dependence structure between the hypotheses. 
In this procedure decisions regarding different hypotheses deterministically depend upon each other and hence the method is referred to as the non-marginal procedure. The decision rule also 
has the desirable oracle property corresponding to an additive ``0-1" loss function (see \ctp{chandra2017} for the details).
In Section \ref{subsec:new_proposal} we discuss the Bayesian non-marginal procedure briefly. 

\subsection{An overview of the new Bayesian procedure for obtaining non-marginal decisions}
\label{subsec:new_proposal}
\subsubsection{The general multiple testing set-up}
\label{subsubsec:basicsetup}

Let $\bZ=(\bZ_1, \ldots,\bZ_n)$ be the observed data. Let the joint distribution of $\bZ$ given $\bpsi=(\psi_1,\ldots,\psi_m)^T$ be $P_{\bZ|\bpsi}(\cdot)$ where $\bpsi$ are the parameters of interest and $\psi_i\in\varTheta_i$ for all $i=1,\ldots,m$. We put a prior $\Pi(\cdot)$ on the parameter space. Let $\postp(\cdot)$ and $\pexp(\cdot)$ be the posterior probability and posterior expectation of $\bpsi$, respectively, given $\bZ$. Here $P_{\bZ}(\cdot)$ and $E_{\bZ}(\cdot)$ 
represent the marginal distribution of $\bZ$ and expectation with respect to this marginal distribution, respectively.

Consider the following hypotheses:
$$ H_{0i}:\psi_i \in \varTheta_{0i}  \hbox{ versus } H_{1i} : \psi_i \in \varTheta_{1i},  $$ 
where $\varTheta_{0i} \bigcap \varTheta_{1i}=\emptyset$ and  $\varTheta_{0i} \bigcup \varTheta_{1i} 
= \varTheta_{i}$, for $i=1,\ldots,m$. In this particular problem $\varTheta_{0i}=[-1,1]$.

Here we discuss the multiple comparison problem in a Bayesian decision theoretic framework, given data $\bZ_n$. 
For $i=1,\ldots,m$, let us first define the following quantities:\\
$d_i=\begin{cases}
1&\text{if the $i^{th}$ hypothesis is rejected;}\\
0&\text{otherwise;}
\end{cases}
\newline 
r_i=\begin{cases}
1&\text{if $H_{1i}$ is true;}\\
0&\text{if $H_{0i}$ is true.} 
\end{cases}$

\subsubsection{New error based criterion}
\label{subsubsec:err}
Let $G_i$ be the set of hypotheses (including hypothesis $i$) whose decisions are highly dependent to that of the 
$i^{th}$ hypothesis. Define the following quantity:\\
$z_i=\begin{cases}
1&\mbox{if $H_{d_j,j}$ is true for all $j\in G_i\setminus\{i\}$;}\\
0&\mbox{otherwise.}
\end{cases}$\\
If $G_i$ is a singleton, then we set $z_i=1$.

Now consider the term 
\begin{equation}
TP=\sum_{i=1}^md_ir_iz_i. 
\label{eq:tp}
\end{equation}
This is the number of cases $i$ for which $d_i=1$, $r_i=1$ and $z_i=1$; in words, $TP$ is the number of cases for which the 
$i^{th}$ decision correctly accepts $H_{1i}$, and all other decisions in $G_i$, which may accept either 
$H_{0j}$ or $H_{1j}$, for $j\neq i$, are correct. We refer to this quantity as the number of \textit{true positives}, 
and maximize its posterior expectation with respect to $\dec$. But there are also errors to be controlled, and \ctp{chandra2017} advocated control of the following error:
\begin{equation}
E=\sum_{i=1}^md_i(1-r_iz_i),
\label{eq:err}
\end{equation}
subject to maximizing $TP$.

Note that $E$ is the total number of cases $i$ for which $d_i=1$, $r_iz_i=0$, that is, 
either the $i^{th}$ hypothesis is wrongly rejected or some other decision(s) in $G_i$ is wrong, or both. This is regarded as the number of \textit{false positives} in our notion.

We will maximize the posterior expectation of $TP$ given by (\ref{eq:tp}) subject to controlling the posterior expectation of $E$. 
%
Hence, with $E$ to be controlled, the function to be maximized is given by
\begin{equation}
f_\beta(\dec)=\pexp\left[\sum_{i=1}^{m}d_i\left(r_iz_i - \beta \right) \right]  =\sum_{i=1}^{m}d_i\left(w_i(\dec) - \beta \right),
\end{equation}
where $w_i(\dec) = \pexp[r_iz_i]= \postp\left(H_{1i}\cap \{\underset{j\neq i, j\in G_i}\cap H_{d_j,j}\}\right)
$ and $\beta$ is a constant lying in $(0,1)$.

\begin{definition}
	Let $\mathbb D$ be the set of all $m$-dimensional binary vectors denoting all possible decision configurations. Define $$\widehat{\bd}=\argmax_{\bd\in\mathbb{D}} f_\beta(\bd),$$ where $0<\beta<1$. Then $\widehat{\bd}$ is the \textit{optimal decision configuration} obtained as the solution of non-marginal multiple testing method.
	\label{def:nmd}
\end{definition}

It is to be noted that the performance of the non-marginal method heavily depends on the choice of $G_1,\ldots,G_m$ since the decisions regarding different hypotheses depend upon each other through the $z_i$ terms defined in Section \ref{subsubsec:err}. \citet{chandra2017} provide detailed discussion regarding this issue, following which the groups have been formed on the basis of prior 
correlation between the parameters. Specifics on group formation are elaborated in \ref{sec:group_choice}. 

Analogous to Type-I and Type-II errors there also exist \textit{false discovery rate} $(FDR)$ and \textit{false non-discovery rate} $(FNR)$ in multiple hypothesis testing. 
\ctp{SanatGhosh08} defined the \textit{posterior} $FDR$ and $FNR$; we broadly adopt these definitions but significantly modify them in our Bayesian multiple testing method to account
for dependence between the hypotheses. Details on multiple testing error measures are 
discussed in \ref{subsubsec:error_rates}.

\section{Application of the Bayesian non-marginal multiple testing method to the miRNA problem}
\label{sec:application}
We apply the non-marginal method to detect miRNAs with statistically significant differential expression in cancer tissues as compared to normal tissues. 
Note that $\beta$ in the definition of the non-marginal procedure is the penalizing factor between Type-I and Type-II error. It balances between the posterior expectations of $TP$ and $E$. 
How much Type-I error is allowed to be incurred depends on the choice of $\beta$. In this application we choose $\beta$ such that $posterior~FDR\approx0.10.$
The estimated $posterior~FNR$ based on the TMCMC samples is 0.04. The discoveries by the non-marginal method are shown in the second column of Table \ref{tab:result} labelled by \textit{NMD}. Out of a total of 522 miRNAs, the non-marginal method has 
identified 12 miRNAs.
\begin{table}[ht]
	\centering
	\caption{List of significantly deregulated miRNAs}
	\label{tab:result}
	\begin{tabular}{|c|c|c|c|c|c|}
		\hline
		miRNA& Method&Deregulation& BF & $\hat{\psi_i}$&95\% CI of $\psi_i$\\ \hline
		hsa-miR-129-2-3p& NMD&  Up& $>100$&-2.11 & $(-2.95 ,-1.35)$  \\
		hsa-miR-548k &NMD & Up & 30.44& -1.77 & $(-2.47, -1.09)$\\
		hsa-miR-622 & NMD& Up & $>100$&  -1.85 & $(-2.57, -1.13)$\\
		hsa-miR-147b & NMD& Up & 2.75 & -1.32 & $(-2.19, -0.56)$ \\
		hsa-miR-124-3p & NMD& Up & 0.32& -1.34 & $(-2.26, -0.45)$\\
		hsa-miR-130b-5p &NMD& Up & 17.41& -1.94 & $(-2.78, -1.10)$\\
		hsa-miR-7-5p & LRBH & Up & 0.03& -0.69 & $(-1.24 , 0.67)$\\
		hsa-miR-31-3p & LRBH & Up &0.10& -0.76 & $(-1.23 , 0.64)$\\
		hsa-miR-31-5p & LRBH & Up &0.06& -0.67 & $(-1.22 , 0.64 )$\\
		hsa-miR-133b & NMD& Down & $>100$&  2.87 & $(2.02 , 3.65)$\\
		hsa-miR-375& NMD& Down &  2.96 & 1.55 & $(0.73 , 2.36)$\\
		hsa-miR-1249 & NMD& Down & $>100$& 3.46 & $(2.61 , 4.33)$\\
		hsa-miR-1&  NMD, LRBH & Down & $>100$ &  3.78 & $(2.96 , 4.62)$\\
		hsa-miR-133a-3p & NMD, LRBH& Down &$>100$& 4.36& $(3.56, 5.22)$\\		
		hsa-miR-206 & NMD, LRBH& Down & $>100$& 4.59 & $(3.49 , 5.51)$\\		
		hsa-miR-204-5p& LRBH & Down & 0.02&  0.41& $(-1.23 , 0.63)$\\\hline
		hsa-miR-1293& \pbox{5cm}{Declared significant by\\\ctp{nde14}}& Up & 0.15& -0.87 & $(-1.24 , 0.70)$\\\hline
	\end{tabular} 
\end{table}

\subsection{Biological significances of the discoveries made by the non-marginal method}
\label{subsec:biological_significances}
Most of the findings obtained by the non-marginal method have biological significance. hsa-miR-129-2-3p is reported earlier to promote chemo-resistance in breast cancer \ctn{mir129},
and hsa-miR-548k often triggers head and neck cancer by modifying \textit{TP53} gene \ctn{548k}.  hsa-miR-147b is reported to trigger head and neck squamous cell carcinoma 
with high statistical significance \ctn{147b}, and hsa-miR-130b-5p is found to be upregulated in more than five cancer types \ctn{130b5p}. One of the target genes of hsa-miR-124-3p is ADIPOR2 which was reported to be negatively associated with tumour progression in prostrate cancer \ctn{hiyoshi2012}.

hsa-miR-375 represses cell viability and proliferation via \textit{SLC7A11} in oral cancer \ctn{375}. The miRNA hsa-miR-1249 is known to regulate tumour growth via positive feedback loop of Hedgehog signalling pathway \ctn{1249}. 
Both hsa-miR-1 and hsa-miR-133a-3p were reported earlier to inhibit cell proliferation and induce apoptosis via \textit{TAGLN2} gene \ctn{nohata2011,Kawakami2012827}.  hsa-miR-206 has been reported earlier as tumour suppressor \ctn{206}. Notably, the last 3 miRNAs mentioned are also found to be significantly downregulated by \ctp{nde14}. hsa-miR-133b is significantly downregulated 
in our analysis and has been reported as tumour suppressor in esophageal and gastric cancer \ctn{133b2,133b}. This was not reported significant by \ctp{nde14}, however, hsa-mir-1, hsa-mir-133a, hsa-mir-206 and hsa-mir-133b were reported to be functionally related in human cancers \ctn{133bfunction}.
This is a clear indication why incorporating the dependence structure in biological data is important and ignoring which is likely to lead to failure to detect important signals.

\subsection{Results of testing the hypotheses using Bayes factor}
\label{subsec:bf}
Apart from applying the non-marginal multiple testing procedure, we also consider testing the hypotheses
using Bayes factor (BF). The BF is a summary of the evidence provided by the data in favour of one scientific theory as compared to another, both represented by statistical models. 
It is often used to compare competing models and Bayesian hypotheses. In multiple hypothesis testing problems, for the $i^{th}$ hypothesis, the marginal Bayes factor is defined as
\begin{align}
B_i=\frac{\postp(\varTheta_{1i})}{\postp(\varTheta_{0i})}\times \frac{\Pi(\varTheta_{0i})}{\Pi(\varTheta_{1i})}, 
\end{align}
that is, the ratio of posterior odds of $H_{1i}$ to its prior odds for all $i=1,\ldots,m$. As summarized by \ctp{bfraftery}, the evidence against $H_{0i}$ that is directed by the 
magnitude of $B_i$ is shown in Table \ref{table:BF}. 
\begin{table}[H]
	\centering
	\caption{Bayes factor summary}
	\label{table:BF}
	\begin{tabular}{|c|c|}
		\hline
		$B_i$& Evidence against $H_{0i}$  \\ \hline
		1 to 3.2 & Not worth more than a bare mention \\ 
		3.2 to 10 & Substantial \\ 
		10 to 100 & Strong \\ 
		$>100$ & Decisive \\\hline
	\end{tabular} 	
\end{table}
The BFs corresponding to the discoveries are shown in the 4-th column of Table \ref{tab:result}. 
We see that for most of the discoveries by the non-marginal method, BF indicates very strong evidence against $H_0$.
However, in two cases the results of the non-marginal method, the BF based results and those reported in the literature do not agree. Details follow.

Although hsa-miR-124-3p has been declared significant by the NMD method, the corresponding BF shows evidence towards $H_0$ being true. Notably, BF provides evidence towards a belief but does not take into account the multiplicity effect. On the other hand, the discoveries by the NMD procedure are the results of proper $FDR$ control. Controlling the $FDR$ at a more conservative level would 
rule out the miRNAs with low BFs as discoveries, however, sometimes at the cost of missing out important signals. 
Hence, instead of such conservative approach, we recommend further biological investigation regarding this discovery.

Previous biological research has reported hsa-miR-622 to be tumour suppressor in esophageal cancer \ctn{622}, though our analysis indicates upregulation. The discrepancy may also be due to different tissue types where hsa-miR-622 have been reported to be downregulated in the literature. However, these findings are particularly interesting and further biological experiments should be conducted with different groups of patients to find their roles in tumorigenic processes.

\section{Comparison of our non-marginal results with those of the BH procedure}
\label{sec:comparison}
Based on an independent normal model, \ctp{nde14} considered the BH multiplicity correction approach to identify the significant miRNAs.
Before we compare their results with ours it is important to discuss some methodological issues associated with computing the $p$-values needed for the BH procedure. 

For the hypothesis testing problem defined in (\ref{eq:hypotheses2}), 
the null hypothesis corresponding to each $\psi_i$ is composite and here even the marginal tests are not straightforward with the frequentist approach. To mitigate this problem, 
\ctp{nde14} computed the sample medians of $Z_{ij};~j=1,\ldots,n$, say $\tilde{Z}_i$. If $\tilde{Z}_i>0$ they tested $H_{0i}:\psi_i\leq1$ versus $H_{1i}:\psi_i>1$, else  they tested 
$H_{0i}:\psi_i\geq-1$ versus $H_{1i}:\psi_i<-1$ for $i=1,\ldots,m$, and obtained the corresponding $p$-values. In this way, each composite test boils down to one-sided 
$t$-test by means of the monotone-likelihood property. 
However, such separate tests based on the sign of the sample medians is not well-justified statistically as the results of such tests may be non-negligibly different
from those of the original composite tests of interest.

As a solution to the above problem, we propose the likelihood-ratio (LR) test for the original composite hypothesis testing problem with respect to the same independent normal model. To obtain the marginal $p$-values we implement a parametric bootstrap method and subsequently apply the BH adjustment for multiplicity correction. We denote this newly implemented method by LRBH in this article. The discoveries are shown in the second column of Table \ref{tab:result} labelled by LRBH. 
The discoveries by the LRBH method are compared with the findings of the non-marginal method. The detailed comparison of the results is provided in Section \ref{subsec:result_comparison}. The $p$-value approximation procedure is discussed in \ref{subsec:lrstat}.

\subsection{Comparison of the results obtained by the LRBH and the non-marginal procedure}
\label{subsec:result_comparison}
For comparability with the non-marginal method whose $FDR$ is controlled at level 0.10, we control the $FDR$ of the LRBH procedure at level 0.10 as well. 
As summarized in Table \ref{tab:result}, out of the 522 miRNAs, the LRBH method has identified 7 miRNAs as significant whereas the non-marginal method has identified 12 miRNAs. 
Only three miRNAs, namely, hsa-miR-1, hsa-miR-133a-3p and hsa-miR-206 turned out to be the common discoveries by the LRBH 
and the non-marginal procedure. It is thus important to investigate the reasons for the discrepant findings, which we attempt below. 

\begin{figure}[h]
	\includegraphics[trim={.05in .24in .39in .7in}, clip, totalheight=0.17 \textheight] {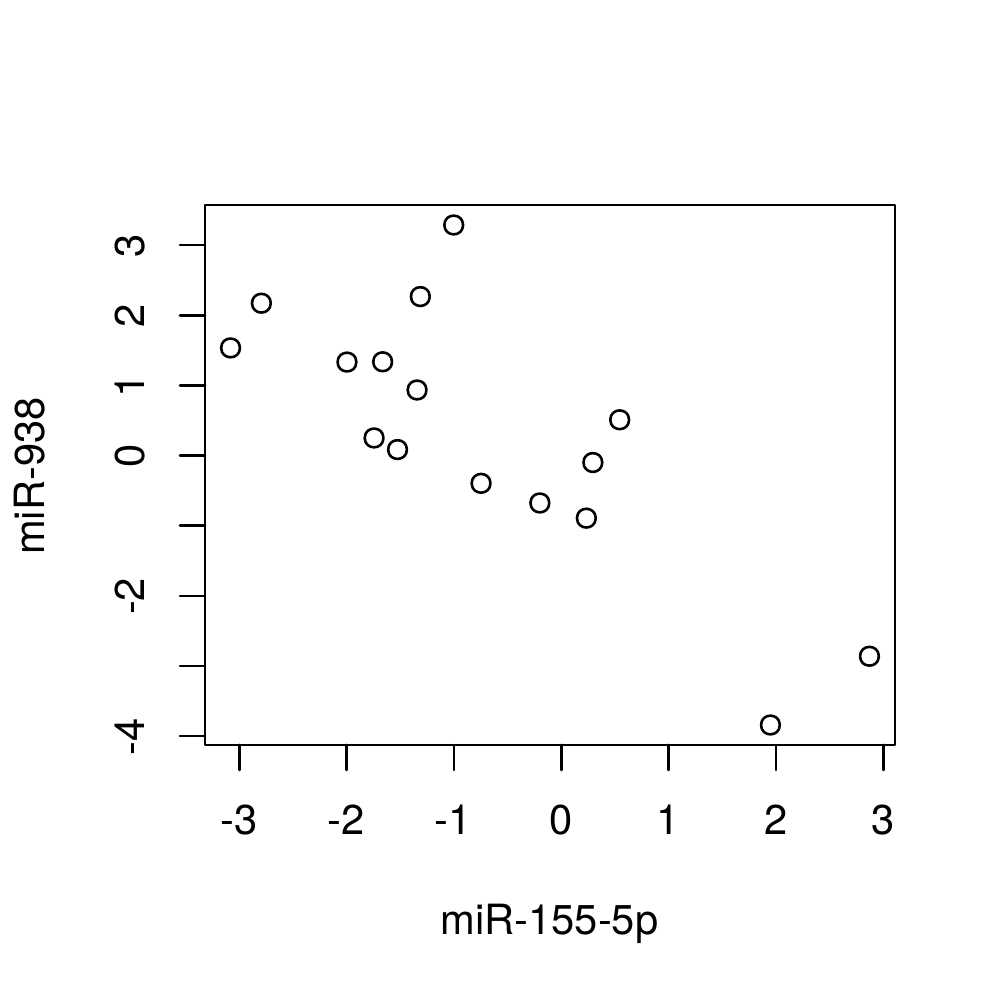}
	\includegraphics[trim={.05in .24in .39in .7in}, clip, totalheight=0.17 \textheight] {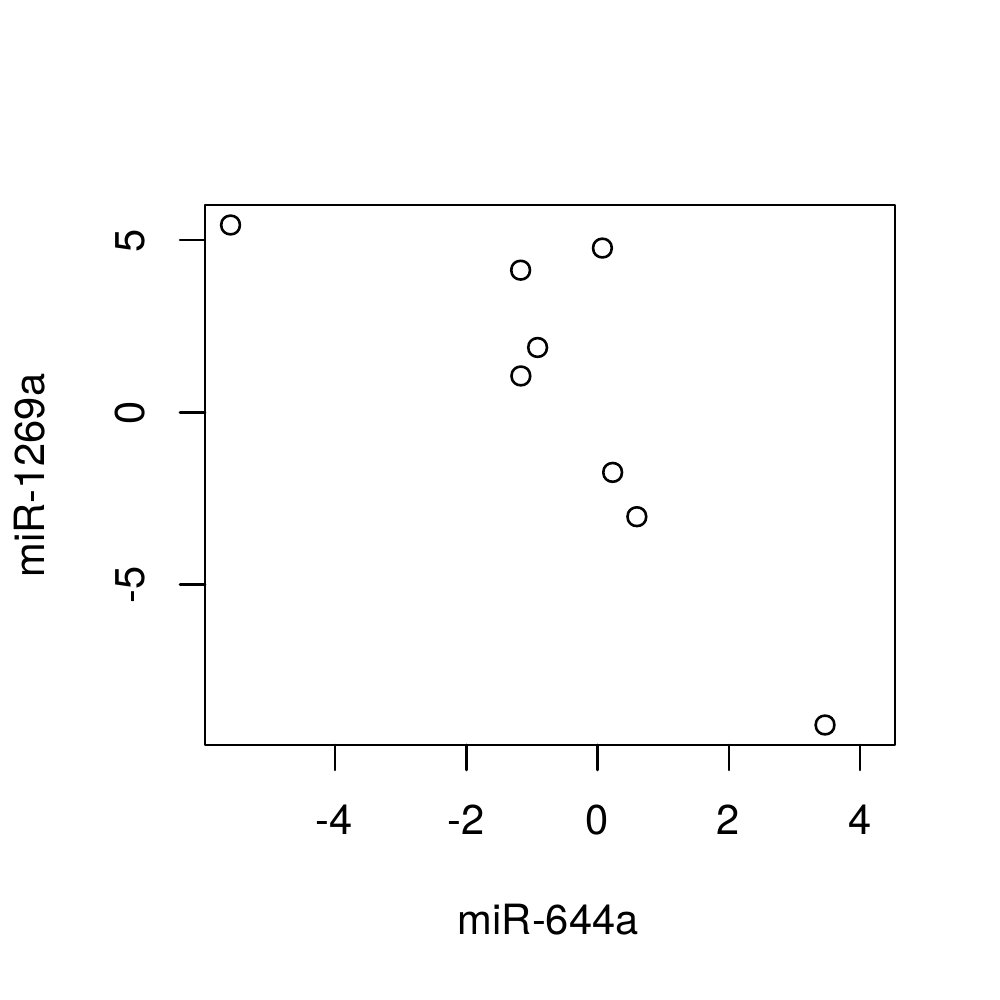}
	\includegraphics[trim={.05in .24in .39in .7in}, clip, totalheight=0.17 \textheight] {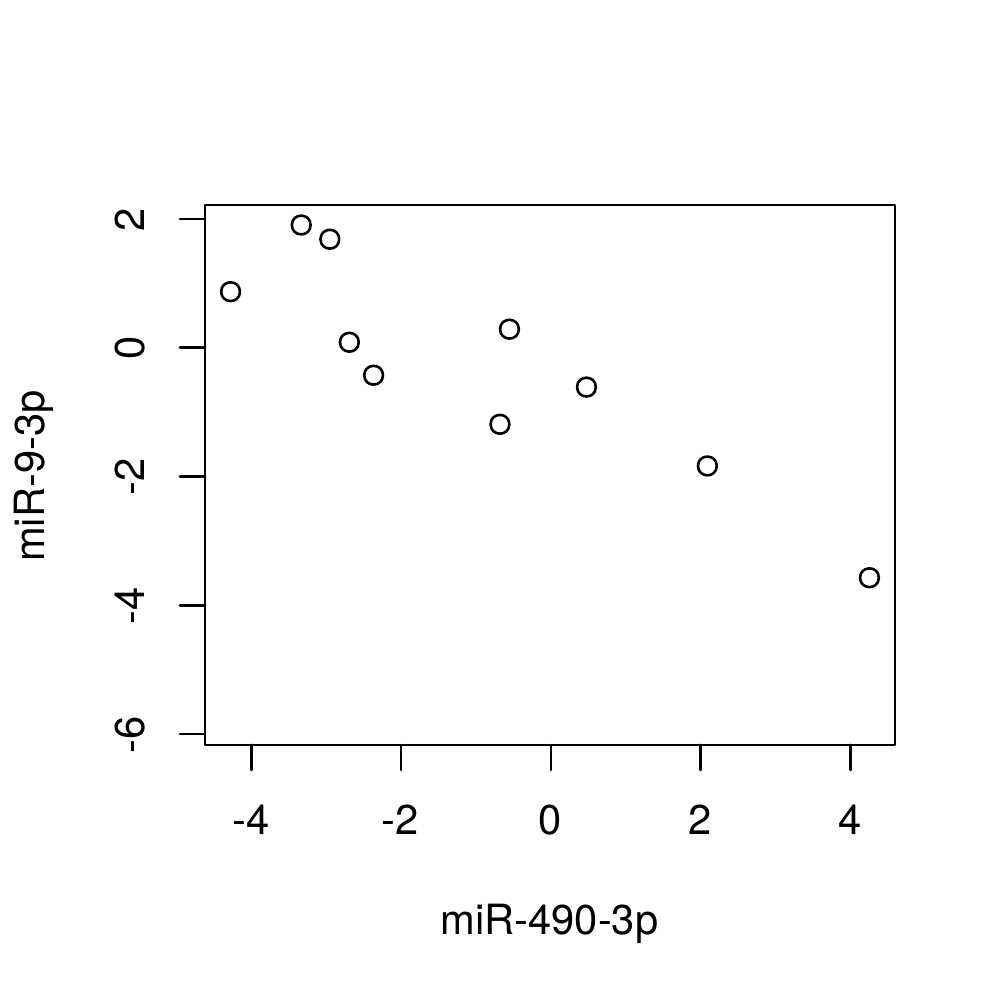}\\
	\includegraphics[trim={.05in .24in .39in .7in}, clip, totalheight=0.17 \textheight] {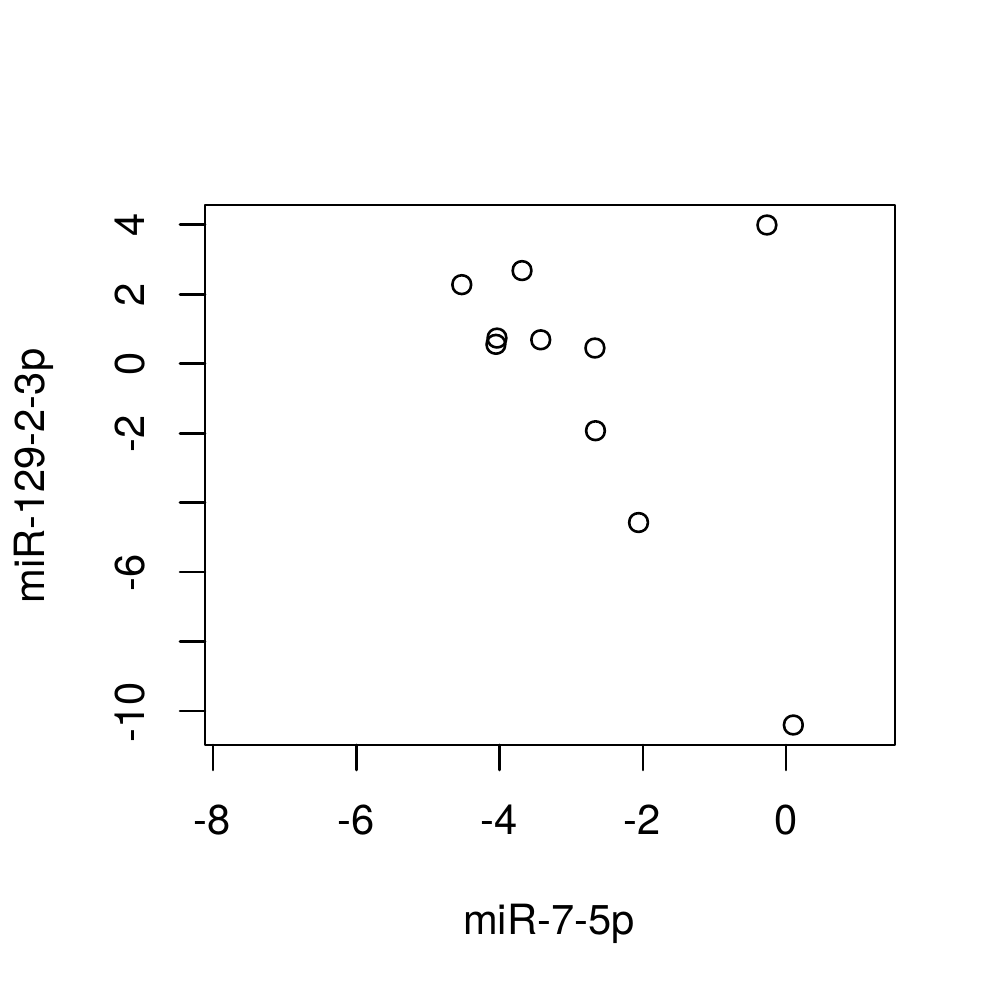}
	\includegraphics[trim={.05in .24in .39in .7in}, clip, totalheight=0.17 \textheight] {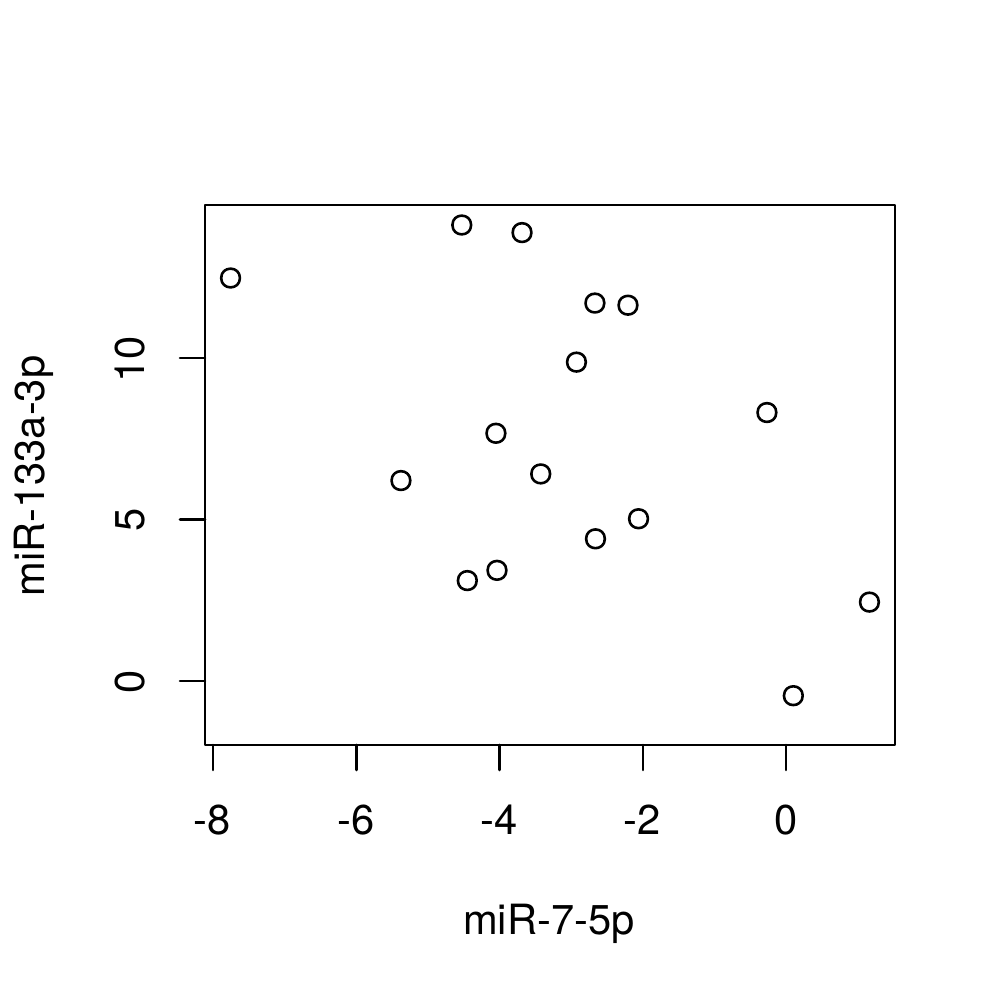}
	\includegraphics[trim={.05in .24in .39in .7in}, clip, totalheight=0.17 \textheight] {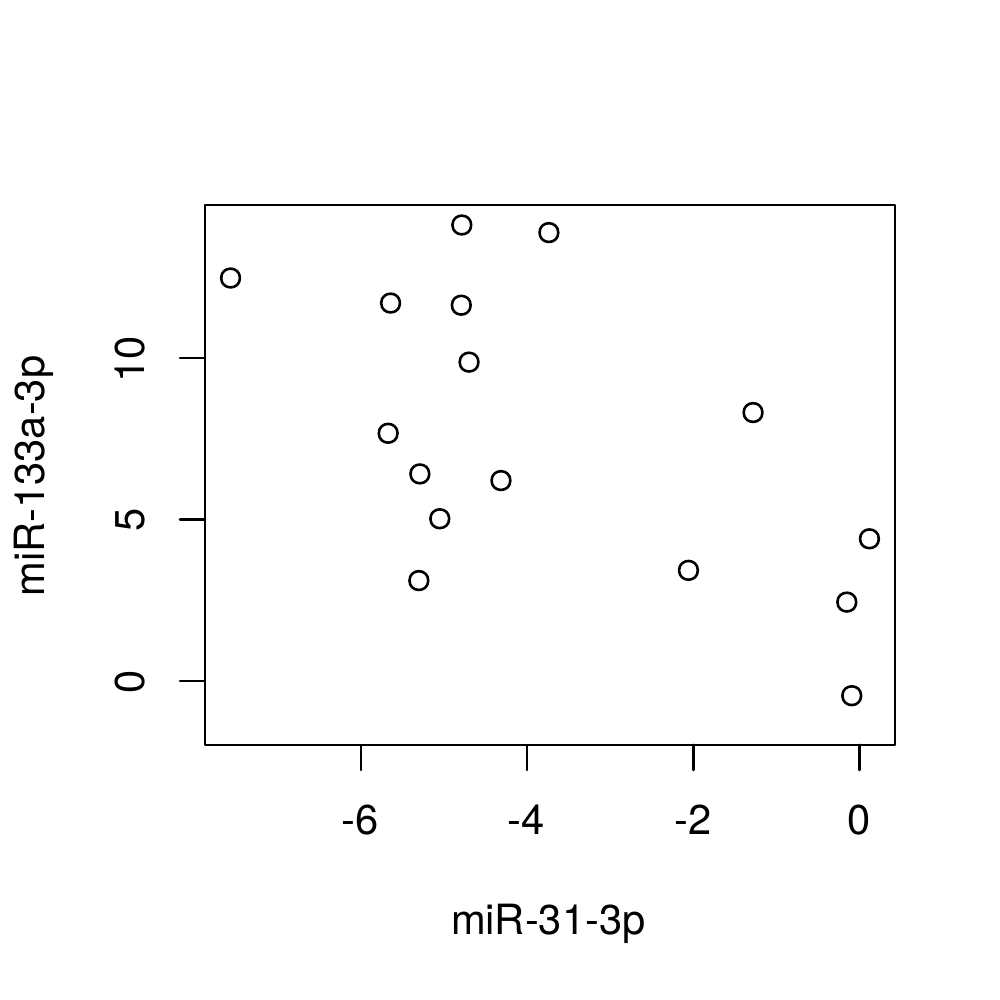}\\
	\includegraphics[trim={.05in .24in .39in .7in}, clip, totalheight=0.17 \textheight] {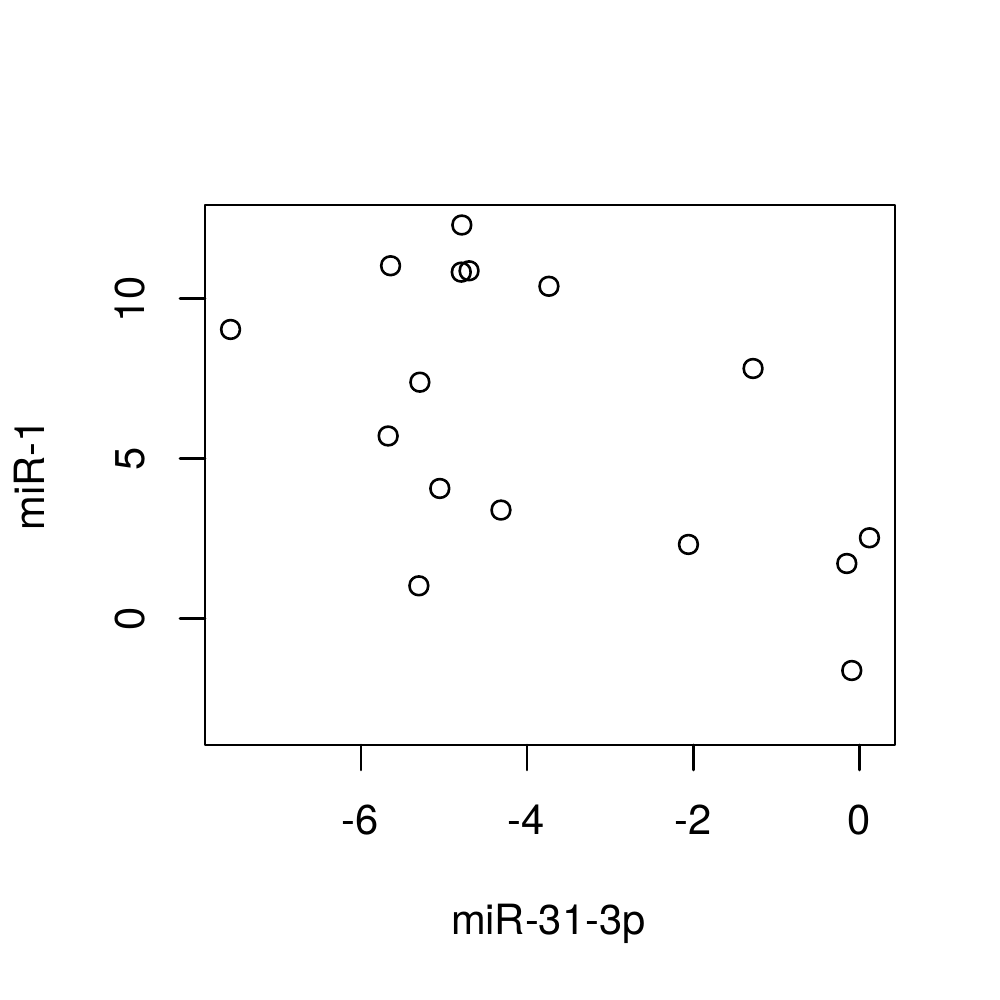}
	\includegraphics[trim={.05in .24in .39in .7in}, clip, totalheight=0.17 \textheight] {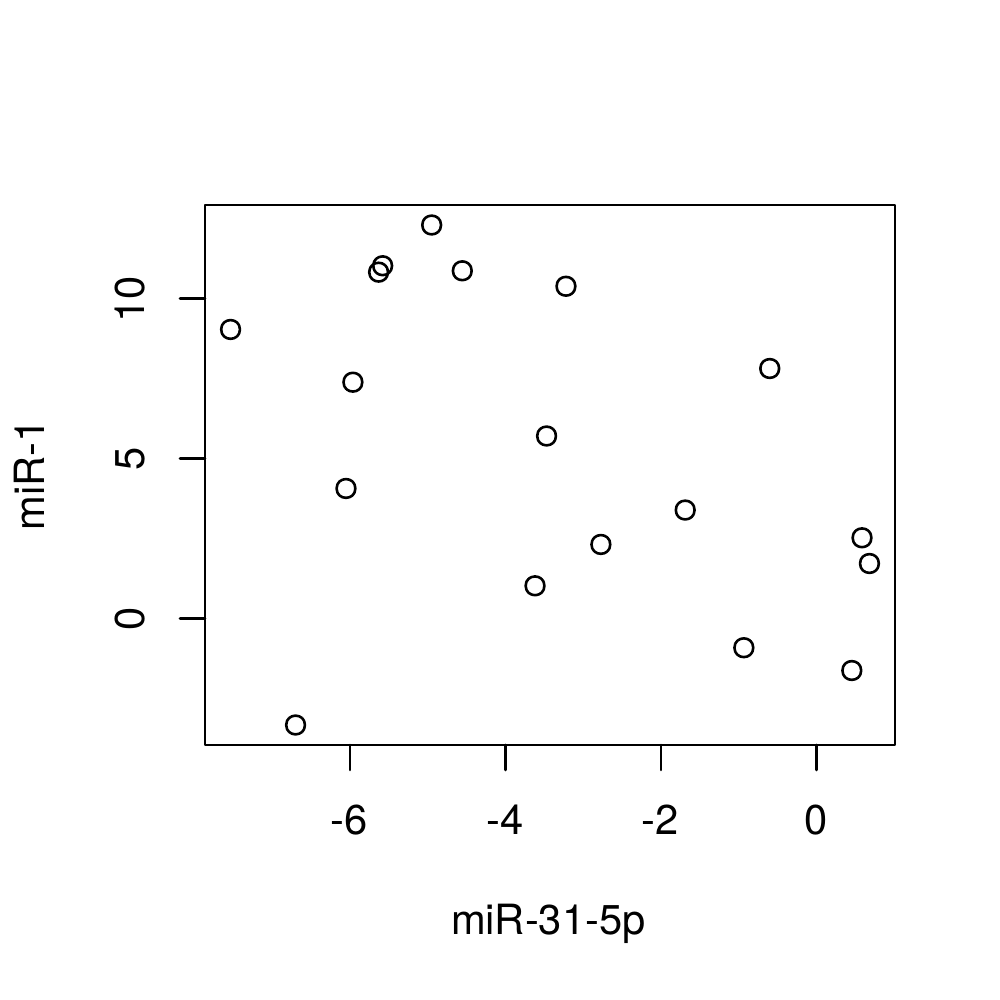}
	\includegraphics[trim={.05in .24in .39in .7in}, clip, totalheight=0.17 \textheight] {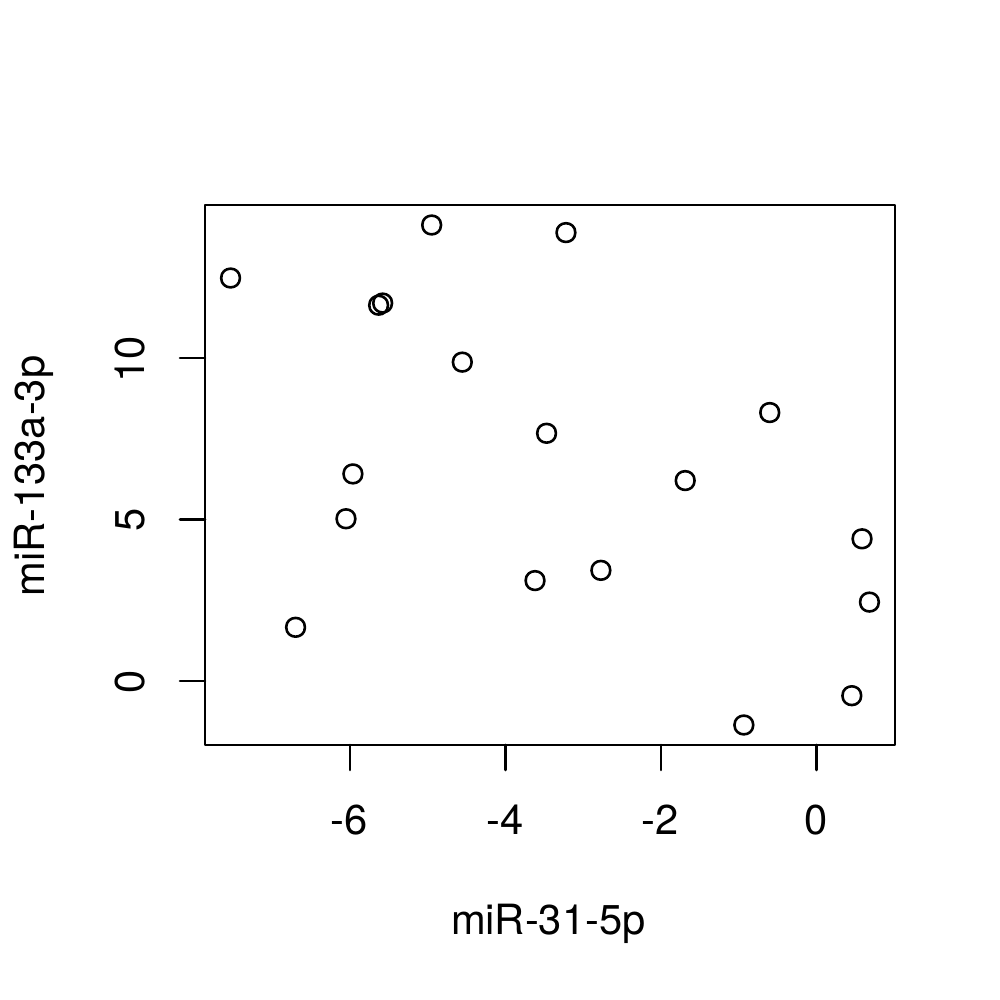}
	\caption{Data exhibiting negative correlations among miRNA expressions.}
	\label{fig:negcorr}
\end{figure}

Note that the expression levels corresponding to different miRNAs including some of the discrepant discoveries exhibit negative correlations (see Figures \ref{fig:heatmap} and \ref{fig:negcorr}). 
In case of negatively correlated test statistics 
validity of general BH procedure of multiplicity correction is not guaranteed theoretically. Also from Section \ref{subsec:lrstat} we see that no dependence between the test statistics has been considered and 
working with only marginal $p$-values practically omits the correlation between the test statistics in the analysis. 
Neglecting the correlation between test statistics might often leads to unstable inference \ctn{qiu2005}.

However, in our Bayesian model, information across miRNAs are pooled and the dependence structure is exploited by means of hierarchical modelling. Not only that, the non-marginal 
method employed to detect statistically significant miRNAs directly takes into account the dependence structure between dependent hypotheses, and the corresponding decisions 
deterministically depend upon each other through the $z_i$ terms defined in Section \ref{subsubsec:err}. Extensive simulation studies show that this method indeed performs 
better in dependent situations as compared to some popular methods (including BH) and asymptotically minimizes the Kullback-Leibler divergence from the true model \ctn{chandra2017}. 
Also for discoveries which are found significant by the BH method only, the Bayes factors exhibit quite strong evidence towards $H_0$ being true. Hence, we argue that in this application, 
where the miRNAs possess inherent dependence structure, results yielded by the non-marginal method are more reliable as compared to the BH procedure. 

%

\section{Summary and conclusion}
\label{sec:conclusion}
To the best of our knowledge our attempt to constitute a Bayesian hierarchical model that realistically accounts for case-control dependence and dependence among the miRNAs  
using the genomic coordinates via a bivariate Gaussian process, is the first ever in the literature. Also first ever in the literature, is the application 
of our Bayesian non-marginal multiple testing procedure that accounts for the dependence among the hypotheses in a way that the decisions on the hypotheses are dependent 
on each other and discovers miRNAs that hitherto seem to be unexplored. Indeed, the discoveries employing the non-marginal procedure with respect to this data differ significantly from those of the very popular 
BH procedure applied in the context of a simple independent normal model. This vindicates the importance of realistic modelling of the dependence structure and the associated
realistic non-marginal Bayesian multiple testing procedure. The BF based evidences are mostly in keeping with the non-marginal testing results except hsa-miR-124-3p. As earlier discussed, performing stricter test will disregard the discoveries with low BF, but we do not recommend that.



Interestingly, all the 12 discoveries made by our non-marginal method are already flagged as significant elsewhere in the literature. However, hsa-miR-622 turned out to be upregulated by our method even though the literature suggests that both are downregulated. In this case, the BF also supports the non-marginal method.  


The above discussion points to the fact that the results associated with our Bayesian non-marginal method are generally well-supported by the literature and the BF values. In realistic multiple testing situations where the number of hypotheses is much larger compared to the sample size, pooling information across hypotheses exploiting their inherent dependence structure seems crucial for realistic inference. The leave-one-out posterior predictive plots (\ref{fig:cvplot}) suggest that the predictive prowess of our model is excellent even with a very small training set. Hence, our approach of Bayesian hierarchical modelling of miRNA expression data is well-supported. 

Although further biological research is necessary to shed more light on the discrepant findings summarized above, 
from the statistical perspective, incorporation of dependence structure in our Bayesian model and the non-marginal testing method is the major cause for the discrepant discoveries.  
Indeed, in our experience, taking account of the underlying dependence structure in the model and the testing method is absolutely necessary for realistic inference in complex phenomena,
as here.

\section{Code with Example and Instructions}
The codes are available in the following link\\\url{https://goo.gl/N9AAbp}.

\begin{center}
{\huge \bf Supplementary Material}
\end{center}

\renewcommand\thefigure{S-\arabic{figure}}
\renewcommand\thetable{S-\arabic{table}}
\renewcommand\thesection{S-\arabic{section}}
\renewcommand{\theequation}{S-\arabic{equation}}
\section{Details on bivariate Gaussian process}
\label{sec:bgp}
We now elaborately explain the latent process and its distribution. For any arbitrary set of $r$ miRNAs in the $i^{th}$ strand, let $\bTheta^{(i)}=\begin{bmatrix}\theta^{(i)}(x_1),&\cdots,&\theta^{(i)}(x_r)\\
\widetilde{\theta}^{(i)}(x_1),&\cdots,&\widetilde{\theta}^{(i)}(x_r)\end{bmatrix}$ be the $2\times k$ matrix of their expression levels, the rows of which denote the expression values 
corresponding to case and control. Then $\bTheta^{(i)}$ follows a \textit{matrix-variate normal} $(\mvn)$ distribution: 
\begin{align*}
\bTheta^{(i)}
\sim \mathcal{MVN}_{2\times r}\left(M^{(i)} (\bx_{1:r}),U^{(i)2\times2},V^{(i)r\times k}(\bx_{1:r} ) \right),
\end{align*}	 
where
\begin{align}
M^{(i)} (\bx_{1:r})&=
\begin{bmatrix}
\mu^{(i)}(x_1),&\cdots,&\mu^{(i)}(x_r)\\
\widetilde{\mu}^{(i)}(x_1),&\cdots,&\widetilde{\mu}^{(i)}(x_r)
\end{bmatrix};\label{eq:meanfunc}\\
V^{(i)} (\bx_{1:r})&= ((c^{(i)}(x_j,x_l ) ))_{r\times r}\label{eq:matsig1}.
\end{align}
In the above, $(x_1,\ldots,x_k)$ are the coordinates of the miRNA genes in the genome.

Let $d$ be the distance between two locations $x_j$ and $x_l$, say. Then, the Mat\'ern covariance function is given by 
\begin{align}
c^{(i)}(x_j,x_j )&={\sigma^{(i)}}^2;\\
c^{(i)}(x_j,x_l )&= {\sigma^{(i)}}^2\frac{2^{1-\nu^{(i)}}}{\Gamma(\nu^{(i)})} \left( \sqrt{2\nu^{(i)}}\frac{d}{\rho^{(i)}} \right)^{\nu^{(i)}} 
K_{\nu^{(i)}}\left(\sqrt{2\nu^{(i)}}\frac{d}{\rho^{(i)}} \right), \label{eq:matern}
\end{align}
where $\Gamma(\cdot)$ is the gamma function, $K_{\nu^{(i)}}(\cdot)$ is the modified Bessel function of the second kind; $\sigma^{(i)},\rho^{(i)}$ and $\nu^{(i)}$ are the non-negative parameters of the covariance function. Here ${\sigma^{(i)}}^2$ is the process variance, $\nu^{(i)}$ is the smoothness parameter and $\rho^{(i)}$ is the correlation length. For different strands we allow the hyperparameters to be different and denote the vector of hyperparameters by $\bsigma$, $\brho$ and $\bnu$, respectively. \ctp{Stein99} discussed why the Mat\'ern class of covariance functions is generally recommended in spatial models, following which we adopted the same.

The matrix normal distribution is related to the multivariate normal distribution ($\mathcal{MN}$) in the following way:
\begin{align*}
&\bTheta^{(i)}\sim \mathcal{MVN}_{2\times k}\left(M^{(i)} ,U^{(i)},V^{(i)} \right)\\
\Leftrightarrow&\vecop{\bTheta^{(i)}}\sim\mathcal{MN}\left(\vecop{M^{(i)}},U^{(i)}\otimes V^{(i)}\right),
\end{align*}
where $\otimes$ denotes the Kronecker product and $\vecop{M}$ denotes vectorization of the matrix $M$. In the above, $U^{(i)}$ is the covariance matrix of the columns and $V^{(i)}$ is the 
covariance matrix of the rows of $\bTheta^{(i)}$. Matrix-normal distributions are particularly useful when there are reasons to believe that the vector valued observations are not independent. 
In that situation $U^{(i)}$ takes into account the dependence between the observations. Taking $U^{(i)}$ to be the identity matrix and the rows of $M^{(i)}$ to be identical reduces the matrix
normal realization to $iid$ multivariate-normal realizations. 

It is to be noted that the covariance parameters of matrix-normal distribution   are non-identifiable in the sense that for any scale factor, $s>0$, we have:
\begin{equation}
\mathcal{MVN}_{2\times k}\left(M^{(i)},U^{(i)},V^{(i)}\right)\equiv\mathcal{MVN}_{2\times k}\left(M^{(i)} ,sU^{(i)},\frac{1}{s}V^{(i)} \right).
\end{equation}
A remedy to this identifiability problem is discussed by \ctp{glanz2013}. It is prescribed to consider $U$ and $V$ as correlation matrices with diagonals as 1, that is,
\begin{equation}
U^{(i)}_{jj}=1\text{ and } V^{(i)}_{jj}=1,
\end{equation}	 
and to introduce a positive parameter $\sigma$ to estimate the scale of the overall covariance. After this amendment the re-parametrized model is:
\begin{equation}
\vecop{\mathcal{MVN}_{2\times k}\left(M^{(i)},U^{(i)},V^{(i)}; \sigma^{(i)} \right)}\sim\mathcal{MN}\left(\vecop{M^{(i)}},{\sigma^{(i)}}^2U^{(i)}\otimes V^{(i)}\right).
\end{equation}
Following this we also consider $U^{(i)}$ as a correlation matrix. The parameter $\sigma^{(i)}$ which is already embedded in $V^{(i)}$ as a  Mat\'ern covariance function parameter, 
measures the scale of the covariance.	A rigorous study on matrix-normal distribution and its properties can be found in \ctp{gupta1999}.


\section{Discussion on prior distribution of the miRNA expression levels}
\label{sec:psidist}
\begin{figure}[h]
	\centering
	\includegraphics[scale=.3]{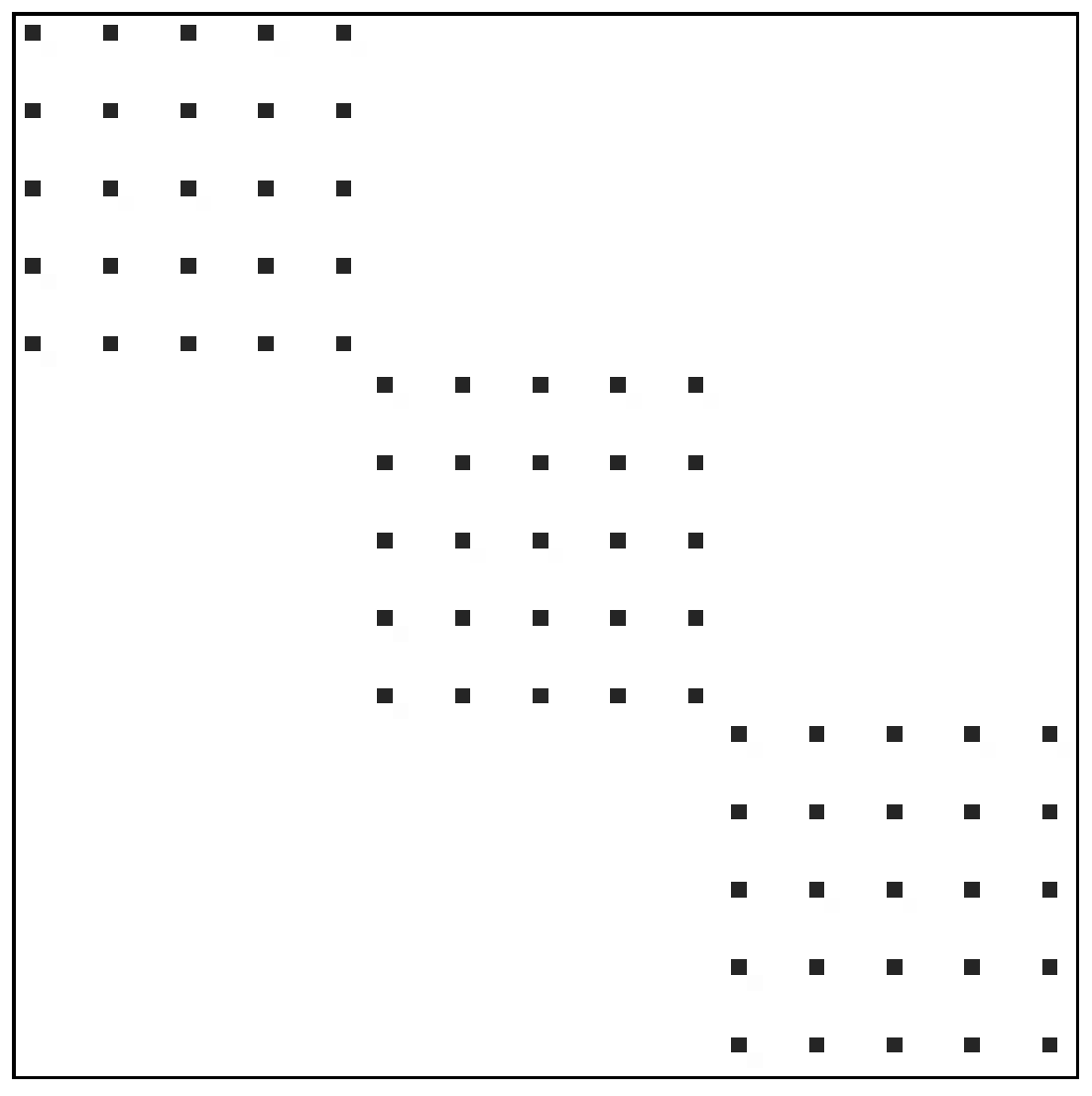}
	\caption{Block-diagonal covariance structure of $\btheta(\bX)$.}
	\label{fig:blockmat}
\end{figure}

The strand-wise bivariate Gaussian process assumption implies that $\btheta(\bx_i)$ has the multivariate normal distribution with 
Mat\'ern covariance given the hyperparameters. Specifically,	
\begin{equation}
\btheta^{(i)}(\bx_i)|\sigma^{(i)},\nu^{(i)},\rho^{(i)} \sim \mathcal{MN}\left(\mu^{(i)}(\bx_i), V^{(i)}(\bx_i) \right).
\end{equation}
Provided a vector of locations $\bx_i$, $\mu^{(i)}(\bx_i)$ and $V^{(i)}(\bx_i)$ represents the mean vector and convariance matrix of $\btheta^{(i)}(\bx_i)$ respectively. Given the hyperparameters, our assumption of independence among the strands makes the distribution of $\btheta(\bX) $ multivariate normal with 
block-diagonal covariance matrix (see Figure \ref{fig:blockmat}).
Indeed, with
\begin{align}
\bmu=&\left(\mu^{(1)}(\bx_1),\cdots, \mu^{(k)}(\bx_k) \right)^T;\\
\bV=&\diag\left(V^{(1)}(\bx_1),\cdots, V^{(k)}(\bx_k) \right)^T,
\end{align}
the distribution of $\btheta(\bX)$ is given by
\begin{equation}
\btheta(\bX)|\bsigma,\bnu,\brho \sim \mathcal{MN}\left(\bmu, \bV \right).
\end{equation}	

As we have considered the same mean function for $\btheta$ and $\widetilde{\btheta}$ \textit{a priori}, the conditional distribution of $\widetilde{\btheta}(\bX)$ given hyperparameters is the following:
\begin{equation}
\widetilde{\btheta}(\bX)|\bsigma,\bnu,\brho \sim \mathcal{MN}\left(\bmu, \bV \right).
\end{equation}

Since $\left(\theta^{(i)}(\cdot),\widetilde{\theta}^{(i)}(\cdot)\right)^T$ are independent bivariate Gaussian process over the strand $i$, it follows that
$\gamma^{(i)}(\cdot)=\theta^{(i)}(\cdot)-\widetilde{\theta}^{(i)}(\cdot)$ are independent Gaussian processes $\mathcal{GP}\left(0,|U^{(i)}|\times c^{(i)}(\cdot,\cdot)\right)$ over the strands. Note that the process variance of $\gamma^{(i)}(x)$ is $|U^{(i)}|\times {\sigma^{(i)}}^2$.
Clearly, an identifiability problem arises here between ${\sigma^{(i)}}^2$ and $|U^{(i)}|$. As a remedy to this problem we consider 
$|U^{(i)}|\times{\sigma^{(i)}}^2$ as a single parameter ${\varrho^{(i)}}^2 $. Let
\begin{equation}
\bgamma(\bX)=\btheta(\bX)- \widetilde{\btheta}(\bX),
\end{equation}
that is, the vector of differential expression levels of the miRNAs across all strands. We define
\begin{align}
W^{(i)}(\bx_l)&=|U^{(i)}|\times V^{(i)}(\bx_l)\text{ and}\\
\bW&=\diag\left(W^{(1)}(\bx_1),\cdots, W^{(k)}(\bx_k) \right)	.	
\end{align}
Note that, similar to $V^{(i)}(\bx_i)$, $W^{(i)}(\bx_i)$ is also a covariance matrix.

Then from Theorem \ref{theorem:gp} we have
\begin{align}
&\bgamma(\bX)|\bvarrho,\bnu,\brho \sim \mn(\bzero, \bW) \\
\Rightarrow~ & \bpsi|\bvarrho,\bnu,\brho\sim \mn(\bzero, \bP \bW \bP^T).~[\because~\bpsi=P\bgamma(\bX)]\label{eq:prior_prob}
\end{align}

\section{Prior distributions on the unknown parameters}
\label{subsec:prior_hyperparameters}
\subsection{Prior on \texorpdfstring{$\bSigma$}{TEXT}}
The Inverse-Wishart distribution is a popular prior on unknown covariance matrices. It is also a conjugate prior for normal likelihoods. Hence, we consider Inverse-Wishart distribution as prior on $\bSigma$  with degrees of freedom $\upsilon$ and parameter-matrix $\delta^2\bI$.
\begin{align}
[\bSigma]&\propto | \delta^2\bI|^{\frac{\upsilon}{2}} |\bSigma|^{-\frac{\upsilon+m+1}{2}}e^{-\frac{1}{2}tr(\delta^2\bI\bSigma^{-1})}\propto 
\delta^{m\upsilon}|\bSigma|^{-\frac{\upsilon+m+1}{2}}\exp\left\{-\frac{\delta^2}{2}tr(\bSigma^{-1})\right\}.
\label{eq:iwprior}
\end{align}
Here $\delta$ is the common scale parameter accounting for the variance of the $Z_{ji}$s. We put inverse-gamma prior on $\delta^2$ and estimate the parameters of the 
inverse-gamma distribution by the empirical Bayes method.  This conjugacy allows $\bSigma$ to be integrated out from the posterior distribution. As $\bSigma$ is a $p\times p$ order matrix $(p=522)$, integrating it out from the posterior density reduces the number of parameters significantly. We have taken 
the degrees of freedom $\upsilon=p+3$ for {\it a priori} second order moment existence.

\subsection{Prior distributions on Mat\'ern hyperparameters }
\begin{figure}[h]
	\centering
	\subfloat[][Prior for $\varrho$.] { \includegraphics[trim={.46in .65in .2in .78in},clip, totalheight=0.165\textheight]{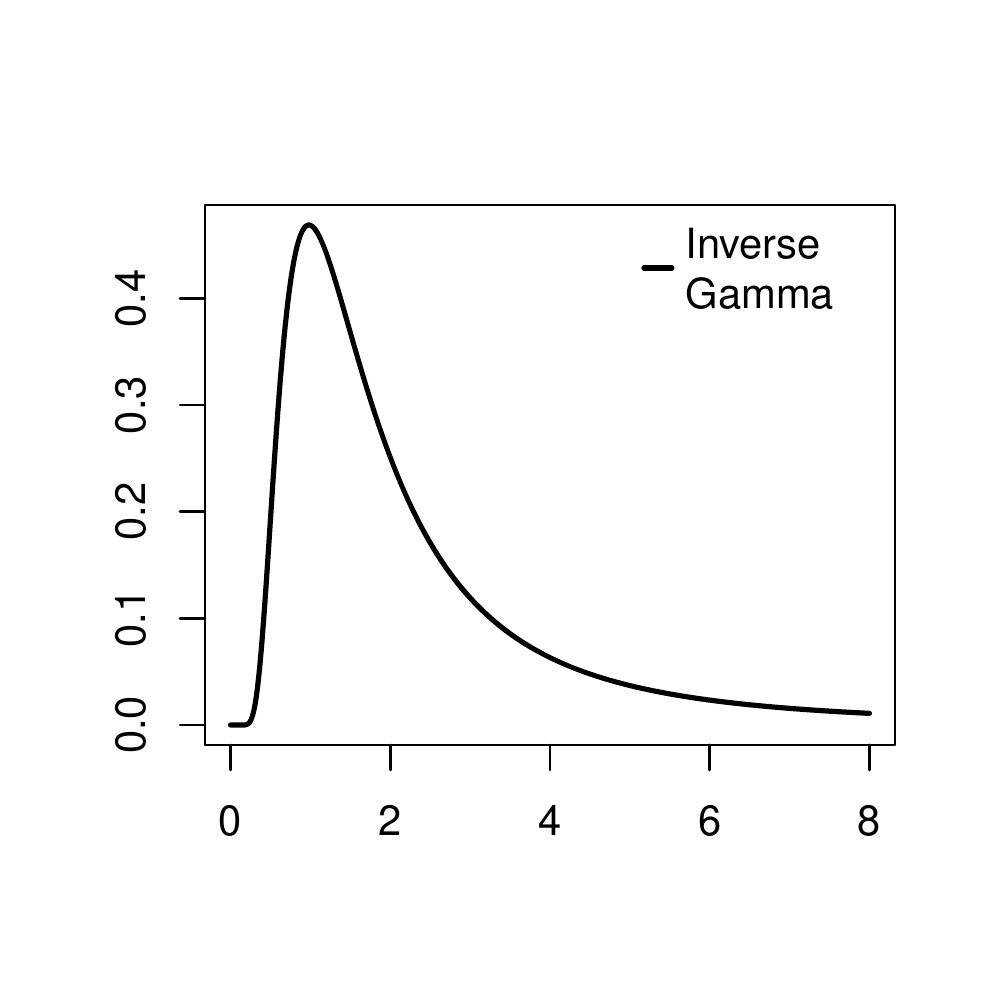}\label{sig}}
	\subfloat[][ Prior for $\nu$.]{ \includegraphics[trim={.46in .65in .2in .78in},clip, totalheight=0.165\textheight]{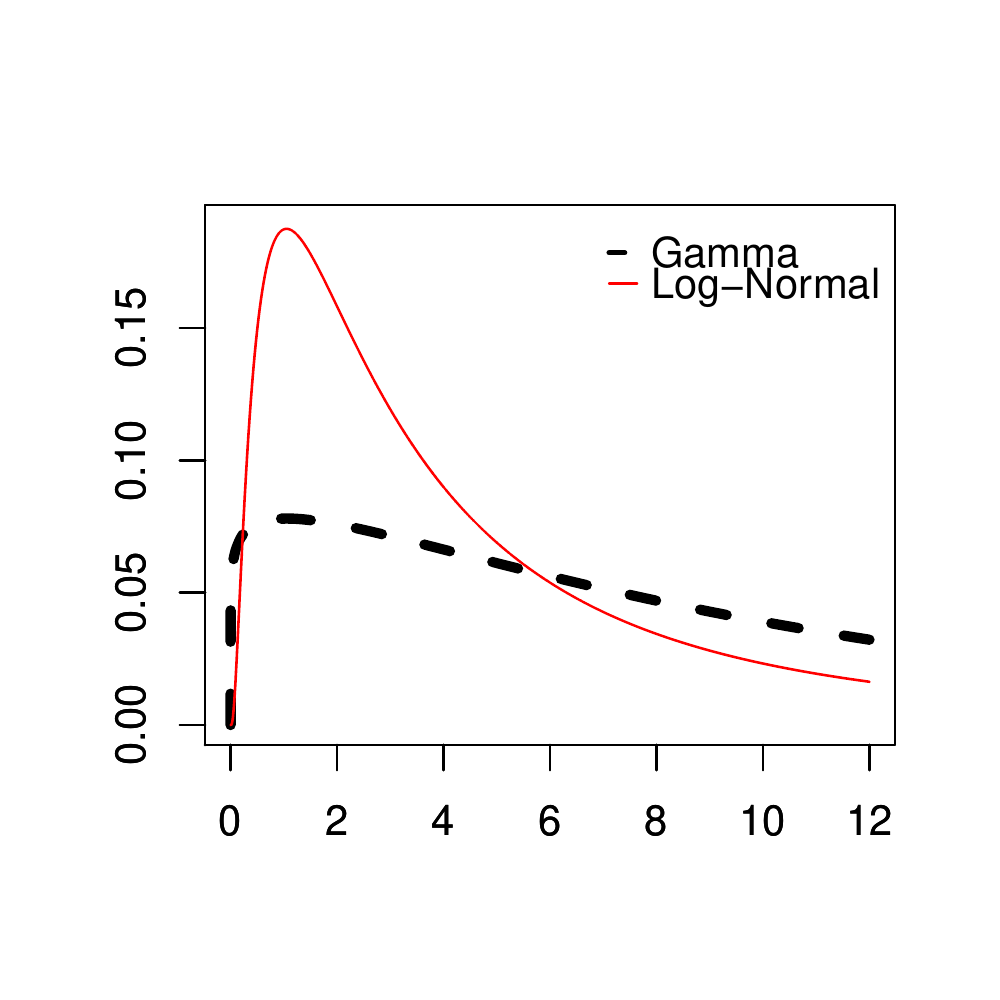}\label{nu}}
	\subfloat[][ Prior for $\rho$.]{\includegraphics[trim={.46in .65in .2in .78in},clip, totalheight=0.165\textheight]{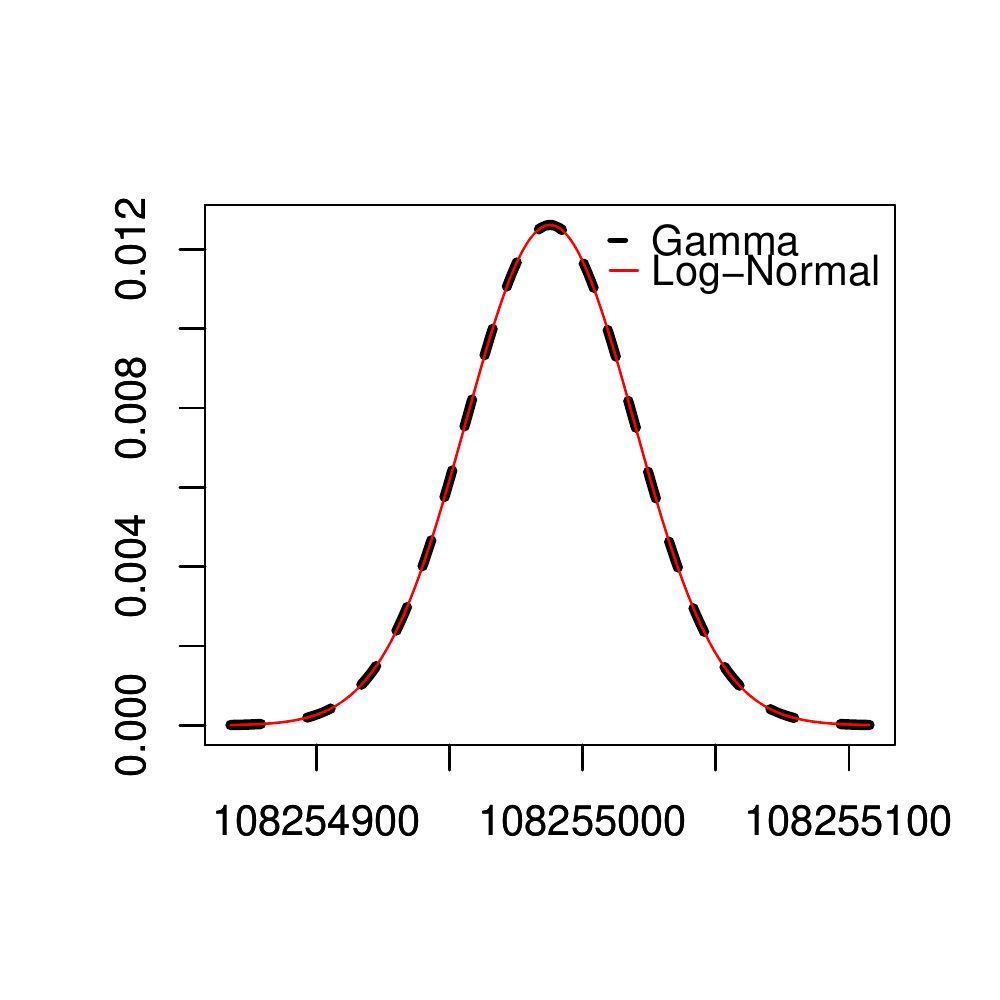}\label{rho}}	\caption{}			
\end{figure}

It is to be noted that $\bvarrho,~\bnu$ and $\brho$ are all unknown positive parameters. As discussed in \ctp{bda2014}, it is desirable that the prior distribution does not unduly 
influence the posterior distribution. As such we put vague prior on these parameters, that is, locally uniform over a widespread range where the true parameter is likely to lie. 
As $\varrho^{(1)},\ldots,\varrho^{(k)}$'s are the strand-wise Gaussian process variances, following the general practice we take the \textit{Inverse-Gamma (IG)} prior on the $\varrho^{(i)}$s. 
The parameters of the IG distribution are adjusted such that the mode of prior distribution is 1 and the variance is 100. This prior gives positive probability to the positive part
of the real line with maximum weight on 1; this modal value can be regarded as the common {\it a priori} summary choice of variance (see Figure \ref{sig}). 

For the prior on $\nu^{(i)}$, note that $\nu^{(i)}$ determines the analytical smoothness of the Gaussian process such that 
$\left(\theta^{(i)}(x),\widetilde{\theta}^{(i)}(x)\right)^T$ is $\lfloor\nu^{(i)}\rfloor$ times mean-square differentiable, where $\lfloor\cdot\rfloor$ is the \textit{floor} function. 
In real data situations not much smoothness is to be expected. This belief is reflected more appropriately by the log-normal prior compared to the gamma prior since the log-normal is 
thin tailed in comparison; see Figure \ref{nu} where we plot gamma and log-normal densities both with mode 1 and variance 100. 
Also observe that the large variance of the log-normal distribution allows even large values of $\nu^{(i)}$ if the data indicates so. As such, we consider the log-normal prior
on $\nu^{(i)}$ with mode 1 and variance 100.

To put a prior on $\rho^{(i)}$ we first need to analyse the role and interpretation of $\rho^{(i)}$ in the model. This parameter is of the dimension of distance \ctn{digglebook} 
and sometimes referred to as the correlation length \ctn{gneiting2010}. In this light, the value of $\rho^{(i)}$ should approximately be the length of the strand over which the observed miRNAs are expressed. We consider the mode of the prior distribution of 
$\rho^{(i)}$s to be the chromosomal length corresponding to strand $i$ with variance 1000. Now the length of genome is really high (generally in order of $10^8$) and with such a large modal value and 
high variance, the gamma and log-normal densities are almost the same (see Figure \ref{rho}). Hence, taking any of the distributions would practically be equivalent and we consider the 
log-normal prior for $\rho^{(i)}$.	

The joint posterior density of $\bpsi,\brho,\bnu,\bvarrho^2,\delta^2$ given data $\bZ$ is:
\begin{align}
&[\bpsi,\brho,\bnu,\bvarrho^2,\delta^2|\bZ]\\ 
\propto& [\bZ| \bpsi,\delta^2]\times [\delta^2] \times  [\bpsi|\brho,\bnu,\bvarrho^2]\times [\brho] \times [\bnu] \times [\bvarrho^2] \\
\propto& \frac{\exp\left\lbrace -\frac{1}{2}\bpsi'(\bP\bW\bP^T)^{-1} \bpsi \right\rbrace  }{\delta^{mn} \left| \bI_n+\frac{1}{\delta^2} (\bZ-\bM) (\bZ-\bM)^T \right|^{\frac{\upsilon+n}{2}} } 
\times [\delta^2] \times [\brho] \times [\bnu] \times [\bvarrho^2] \label{eqn:post_density}
\end{align}
where $\bM=\bpsi^T\otimes \bone_n$. Note that we have integrated out $\bSigma$ from the likelihood. This not only simplifies the likelihood but also reduces the number of parameters 
aiding the MCMC convergence.

\section{Transformation based Markov chain Monte Carlo for sampling from the posterior distribution and leave-one-out cross-validation for model validation}
\label{sec:tmcmc}

\subsection{Additive TMCMC method}
\label{subsec:additive_TMCMC}
TMCMC is employed to generate samples from the joint posterior distribution defined in (\ref{eqn:post_density}). TMCMC is particularly useful for drawing samples from complex high-dimensional distributions. 
We now briefly describe the additive 
TMCMC method, which is what we employ. Suppose one is interested in sampling from the $d$-dimensional distribution with density $\pi(\cdot)$. Let $ g(\cdot)$ be an arbitrary density 
with support $\mathbb{R}_+$, the positive part of the real line. We describe the algorithm in Algorithm \ref{algo:additive_TMCMC}.
\begin{algorithm}
	\caption{Additive TMCMC algorithm}
	\label{algo:additive_TMCMC}
	\begin{algorithmic}[1]
		\State Input: Initial value $\bx_0=(x_{01} , \ldots , x_{0d} )$, number of iterations $N$ and suitably chosen positive valued scaling parameters $c_1,\ldots,c_d$.
		\For{$t=0\cdots N$} 
		\State  Generate $\epsilon\sim g(\cdot)$ and $b_i$, $i=1,\ldots,d$ where $b_i$'s are $iid$ random variables taking values $+1$ or $-1$ with equal probability.
		\State Set $\bx^*=(x_{t1} + b_1c_1 \epsilon,\ldots, x_{td} + b_dc_d \epsilon)$ and $\alpha(\bx_t,\epsilon)=\min\left\lbrace 1, \frac{\pi(\bx^*)}{\pi(\bx_t)} \right\rbrace$.
		\State $\bx_{t+1}=\begin{cases}
		\bx^*\text{ with probability }\alpha(\bx_t,\epsilon)\\
		\bx_t\text{ with probability }1-\alpha(\bx_t,\epsilon).
		\end{cases}$
		\EndFor		
	\end{algorithmic}	
\end{algorithm}

\begin{figure}
	\centering
	\subfloat[][$\psi_{16}$] { \includegraphics[trim={.46in .65in .2in .78in},clip, totalheight=0.175\textheight]{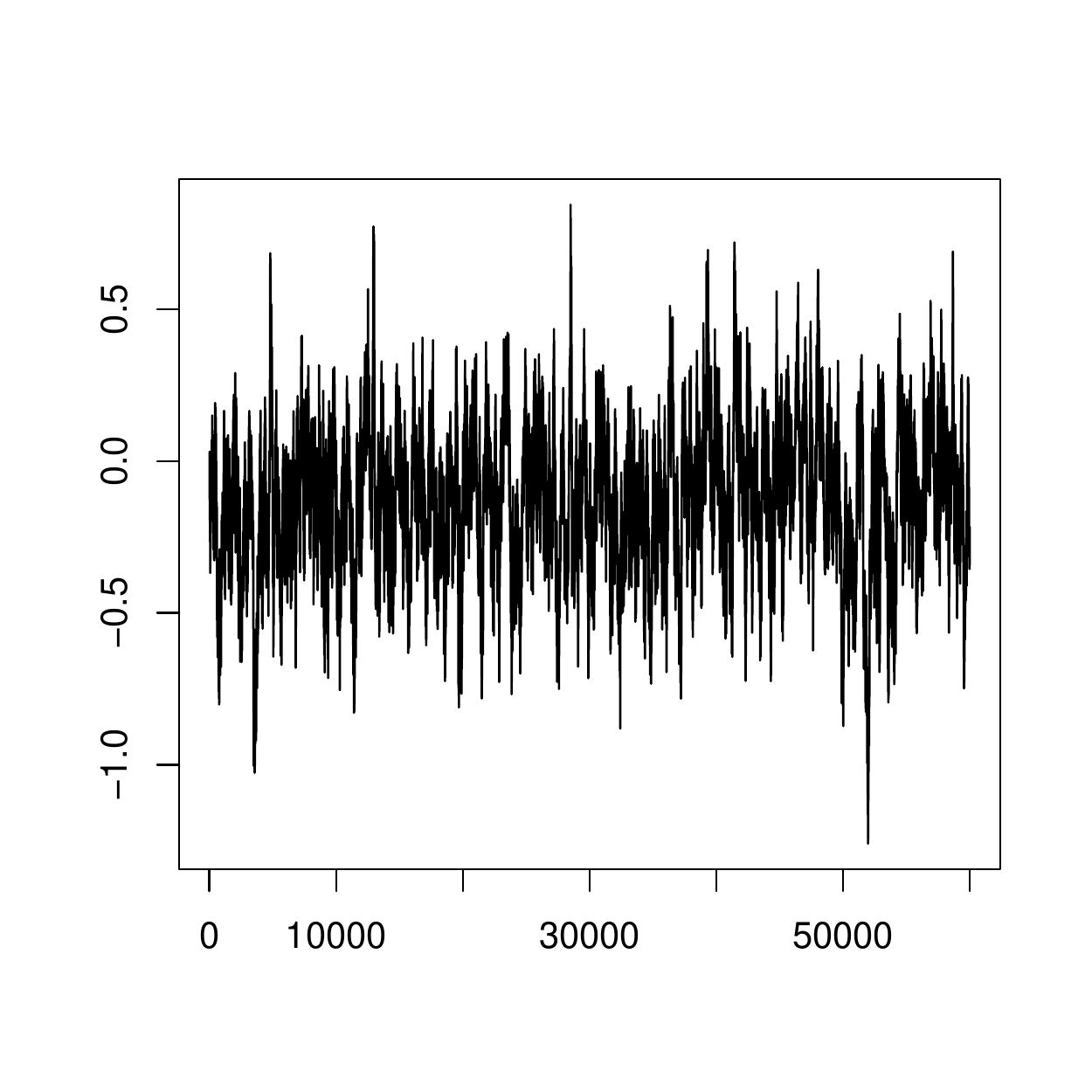}}	
	\subfloat[][$\psi_{254}$]{ \includegraphics[trim={.46in .65in .2in .78in},clip, totalheight=0.175\textheight]{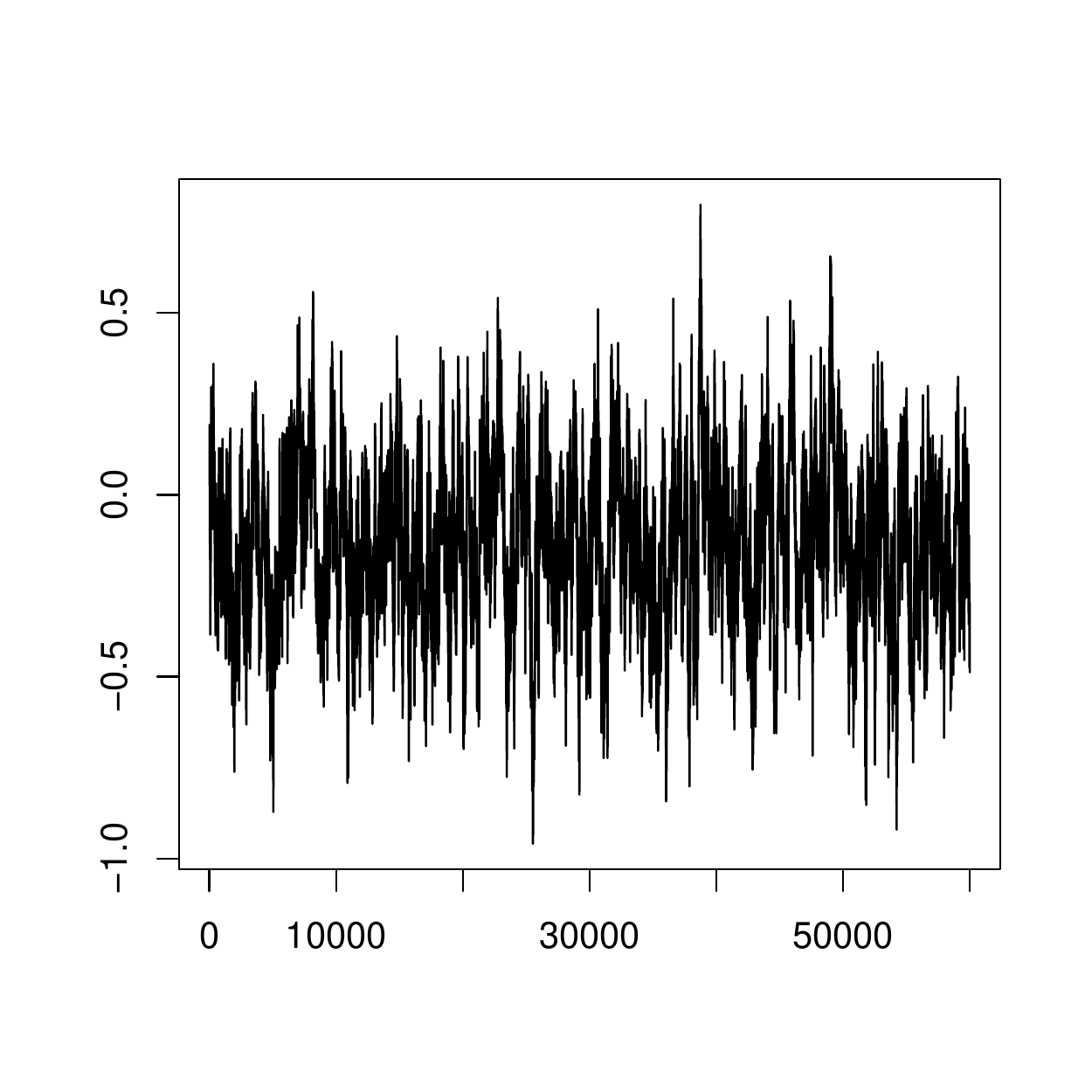}}
	\subfloat[][$\psi_{364}$]{\includegraphics[trim={.46in .65in .2in .78in},clip, totalheight=0.175\textheight]{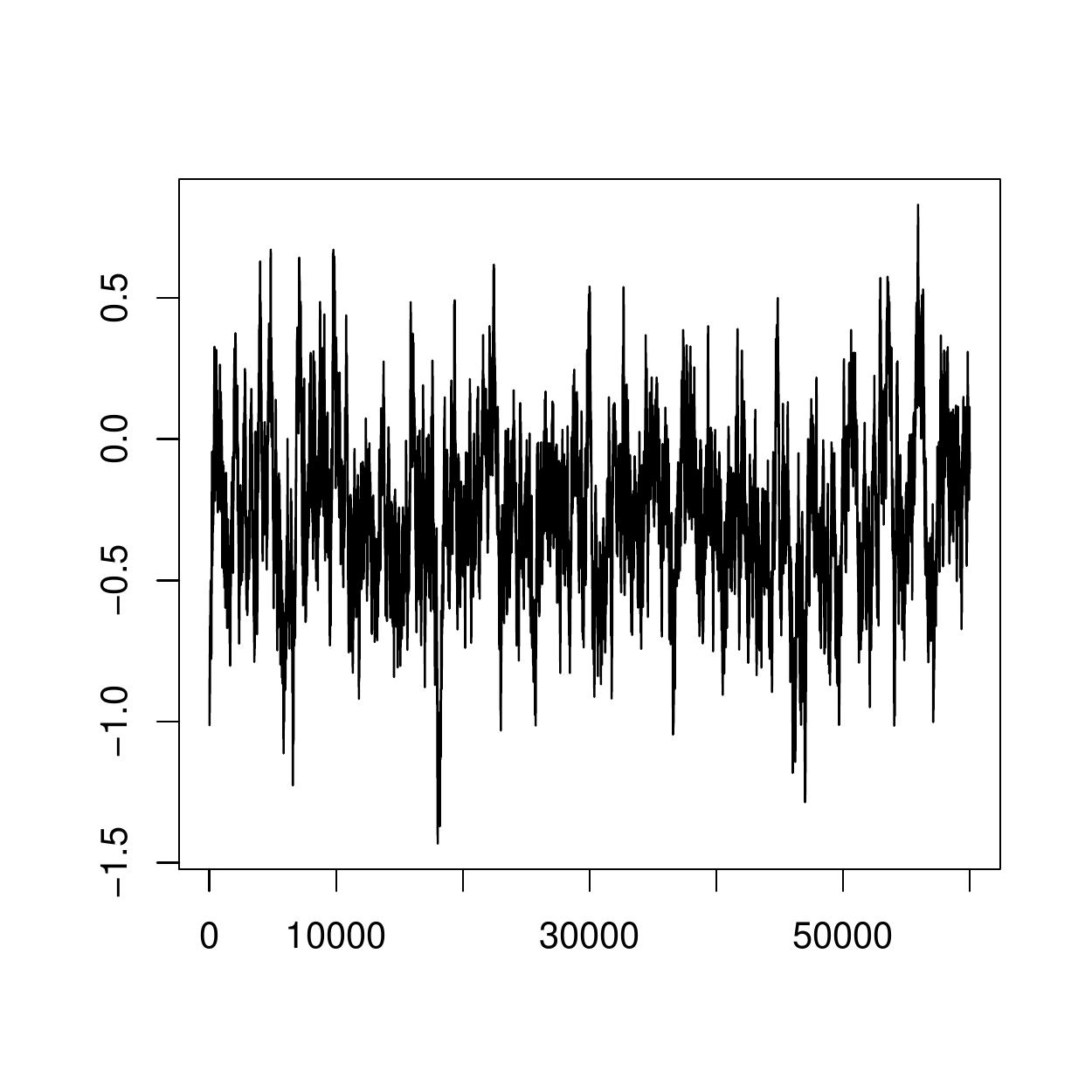}}\\
	\subfloat[][$\log(\varrho_3)$] { \includegraphics[trim={.46in .65in .2in .78in},clip, totalheight=0.175\textheight]{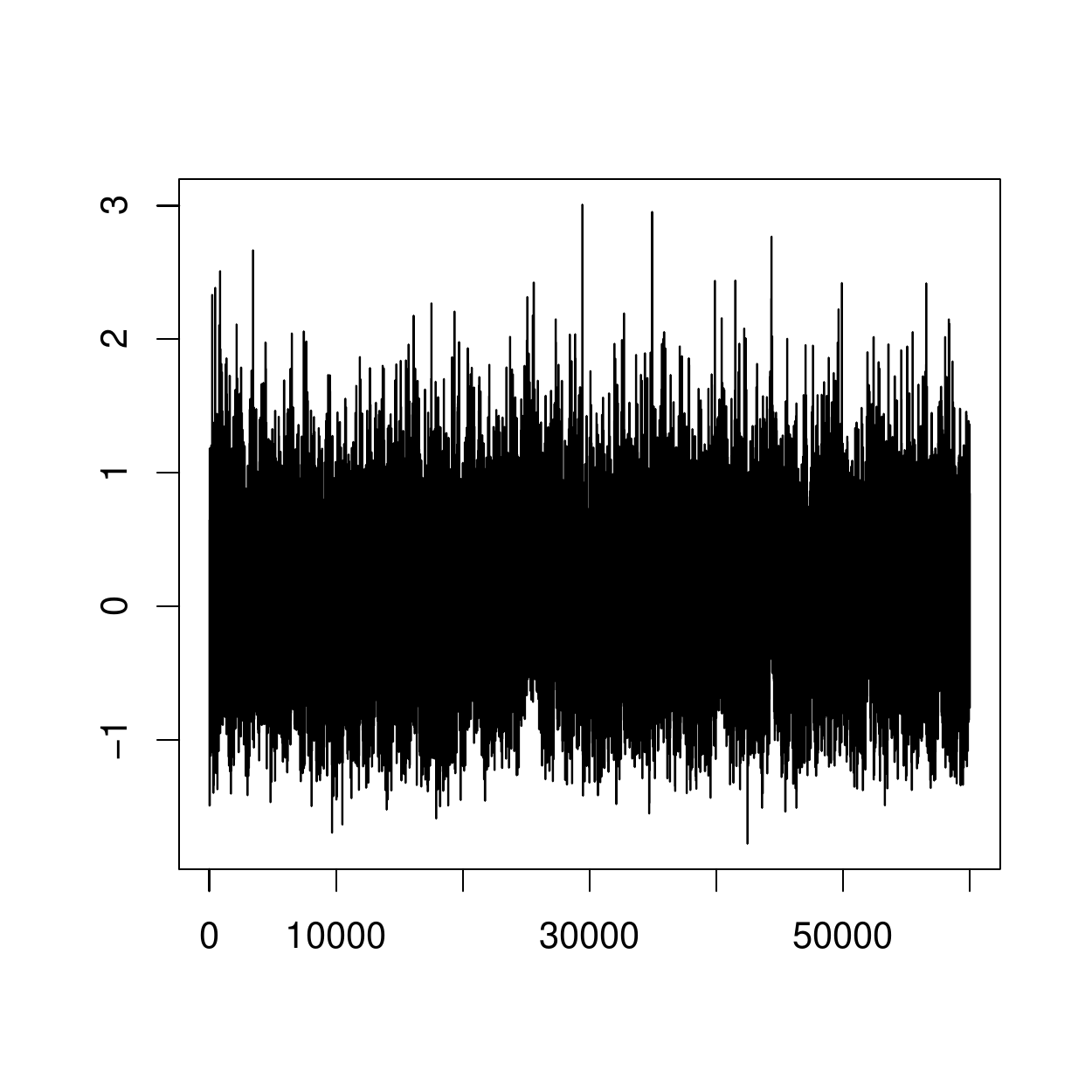}}	
	\subfloat[][$\log(\varrho_{14})$]{ \includegraphics[trim={.46in .65in .2in .78in},clip, totalheight=0.175\textheight]{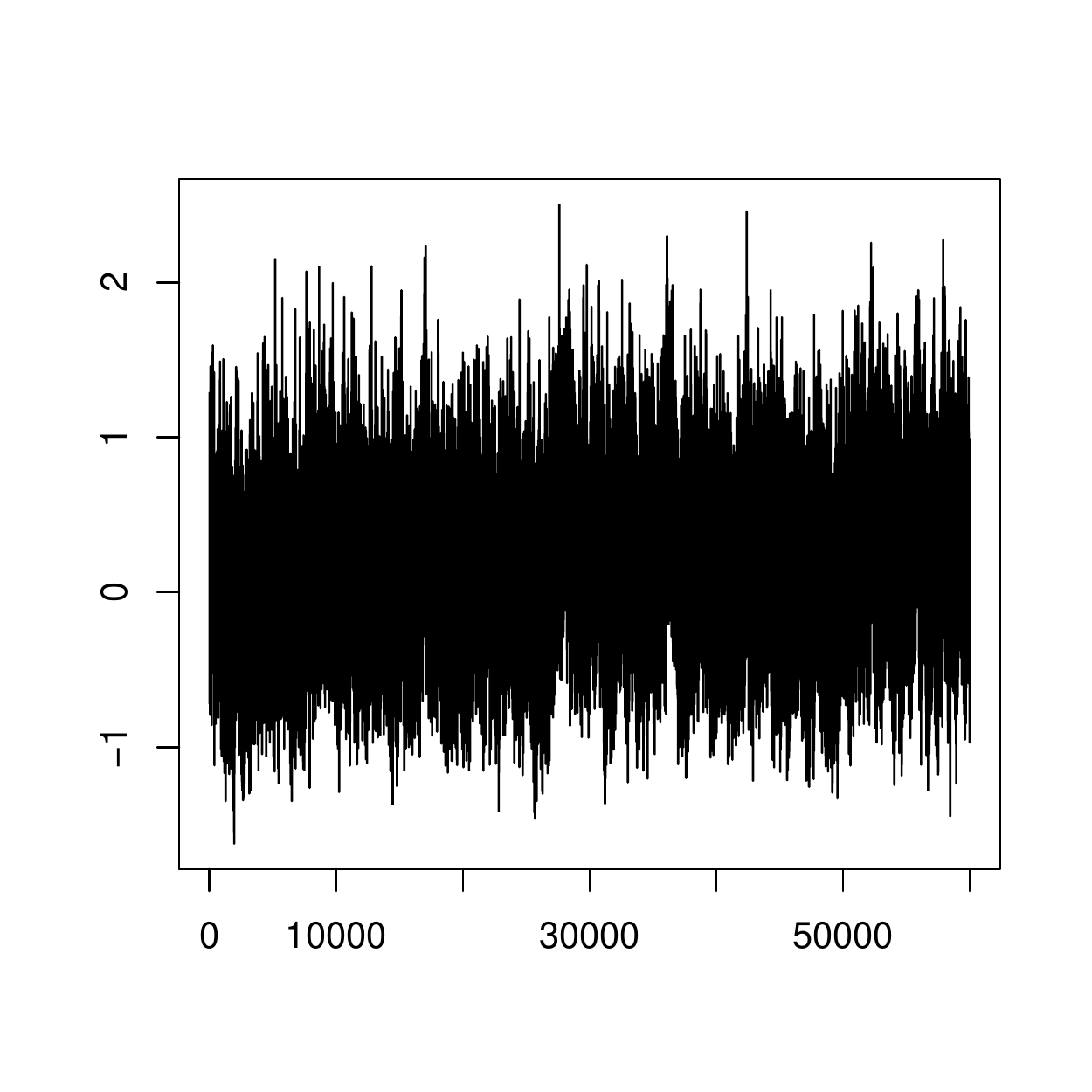}}
	\subfloat[][$\log(\varrho_{36})$]{\includegraphics[trim={.46in .65in .2in .78in},clip, totalheight=0.175\textheight]{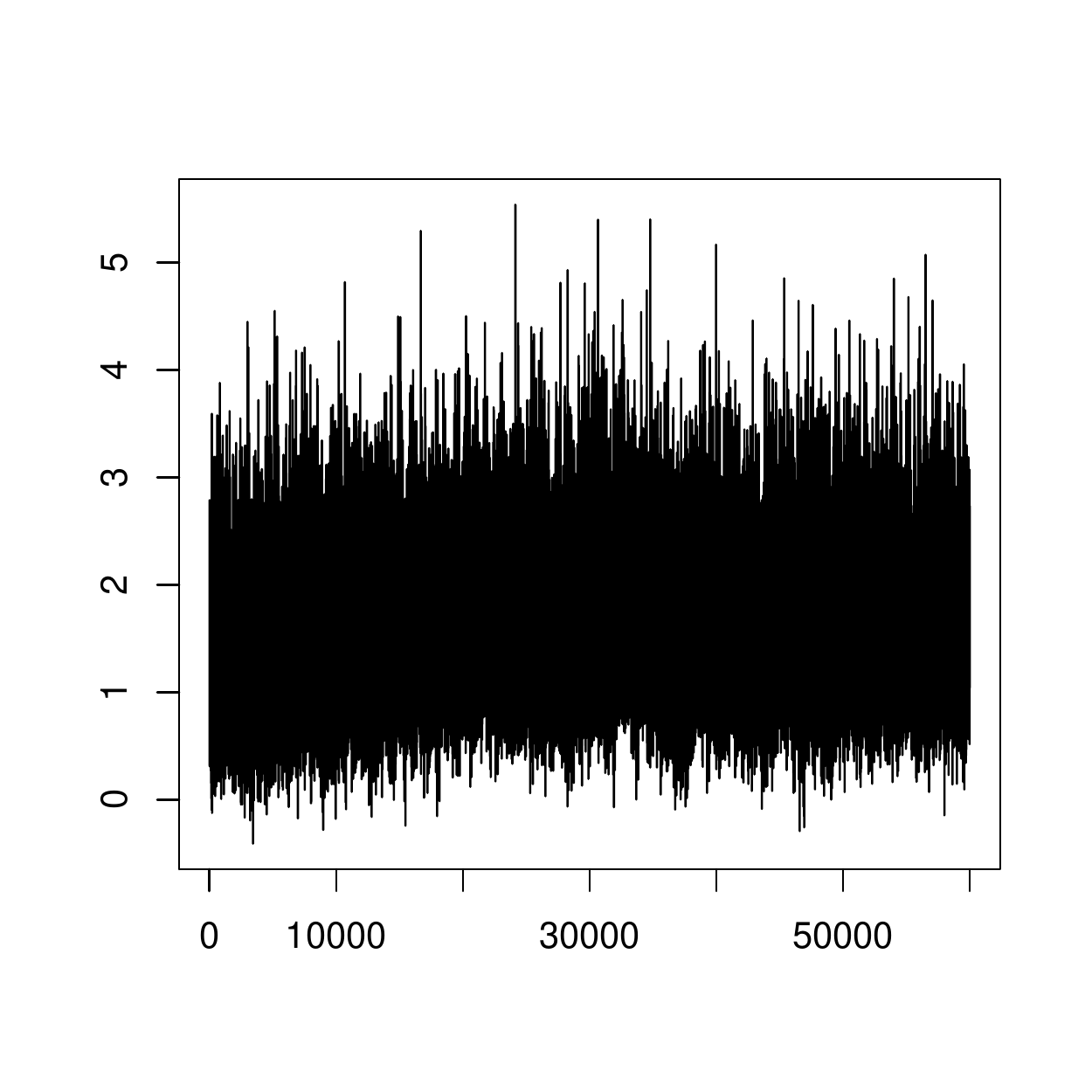}}\\	
	\subfloat[][$\log(\nu_3)$] { \includegraphics[trim={.46in .65in .2in .78in},clip, totalheight=0.175\textheight]{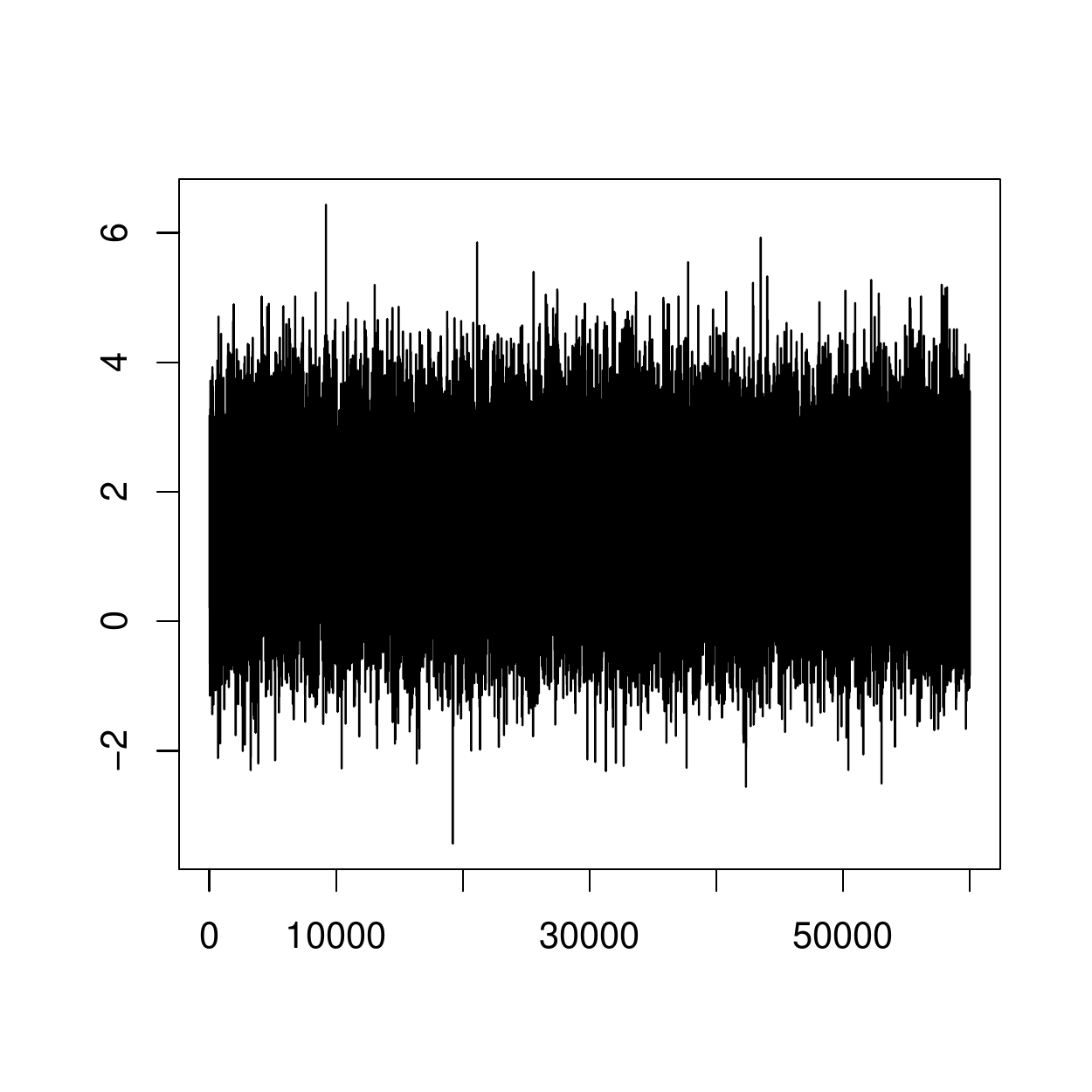}}	
	\subfloat[][$\log(\nu_{18})$]{ \includegraphics[trim={.46in .65in .2in .78in},clip, totalheight=0.175\textheight]{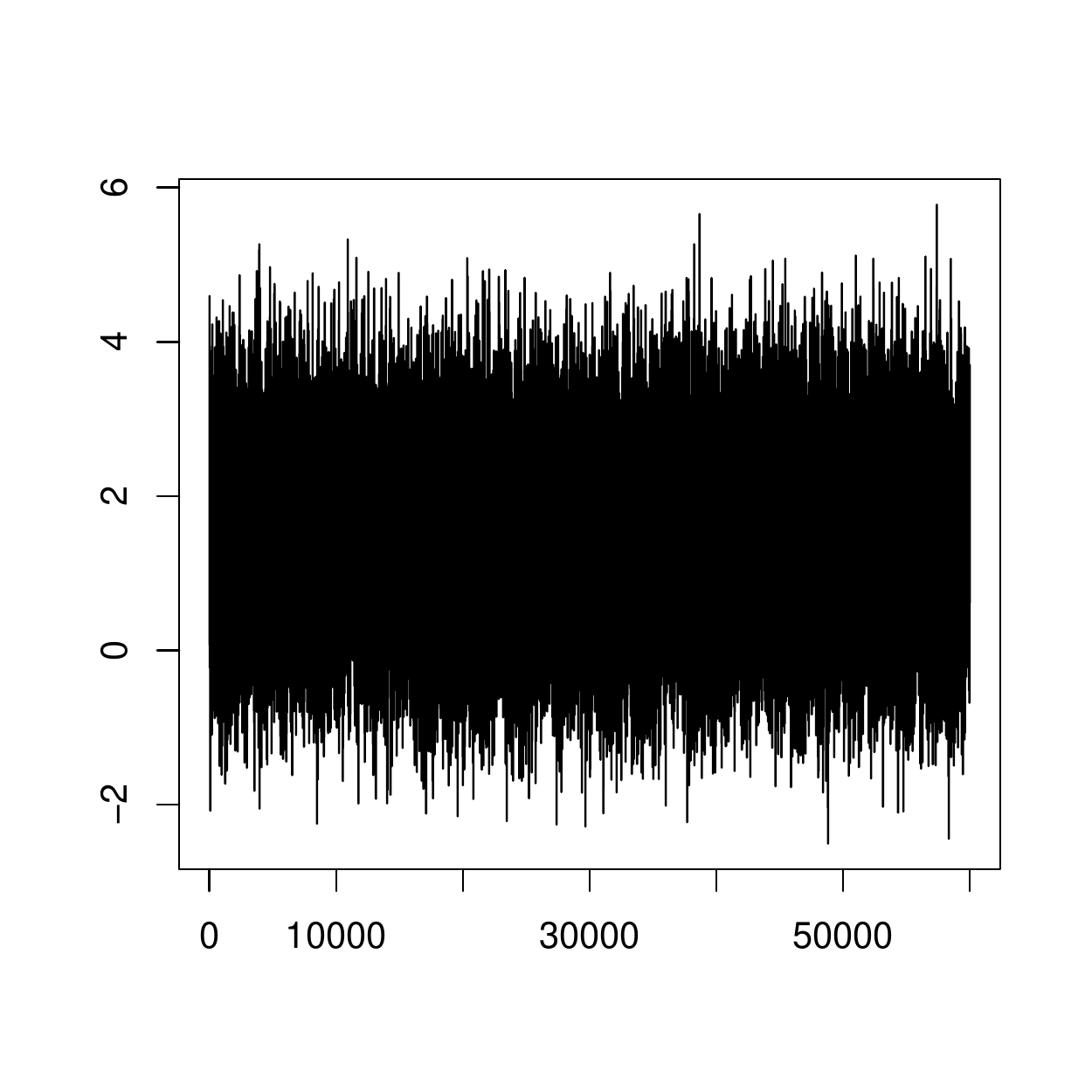}}
	\subfloat[][$\log(\nu_{35})$]{\includegraphics[trim={.46in .65in .2in .78in},clip, totalheight=0.175\textheight]{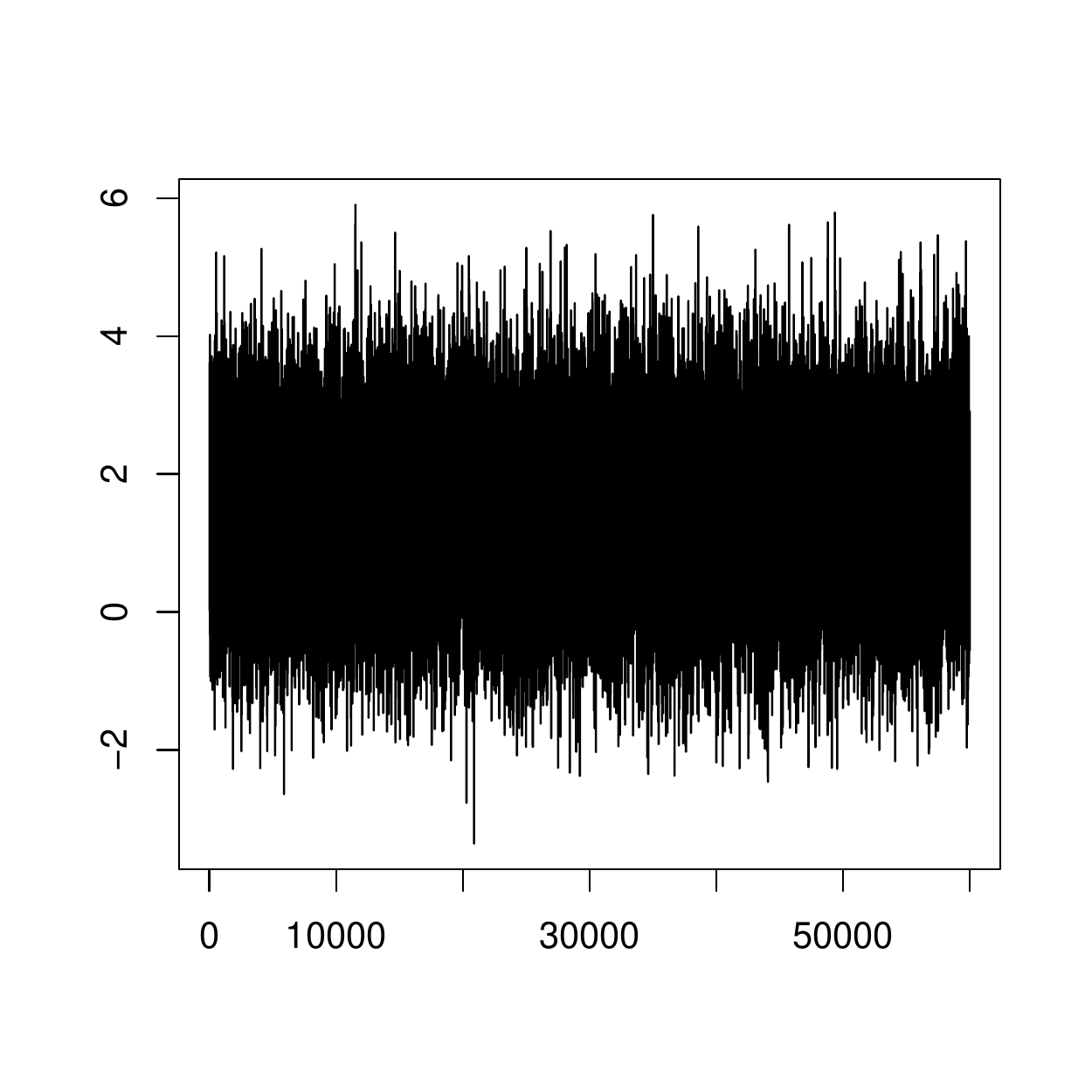}}	\\
	\subfloat[][$\rho_2$] { \includegraphics[trim={.46in .65in .2in .78in},clip, totalheight=0.175\textheight]{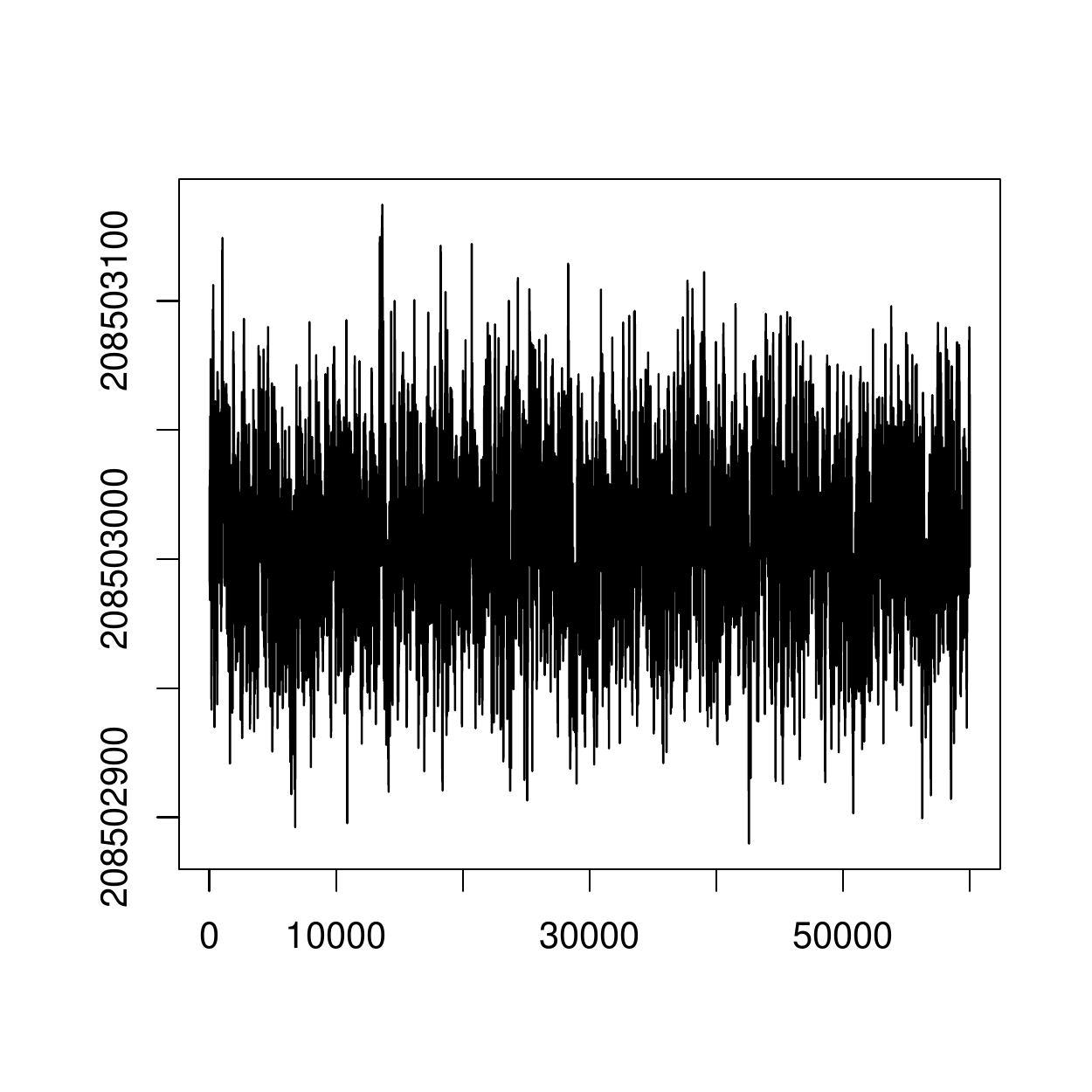}}	
	\subfloat[][$\rho_{19}$]{ \includegraphics[trim={.46in .65in .2in .78in},clip, totalheight=0.175\textheight]{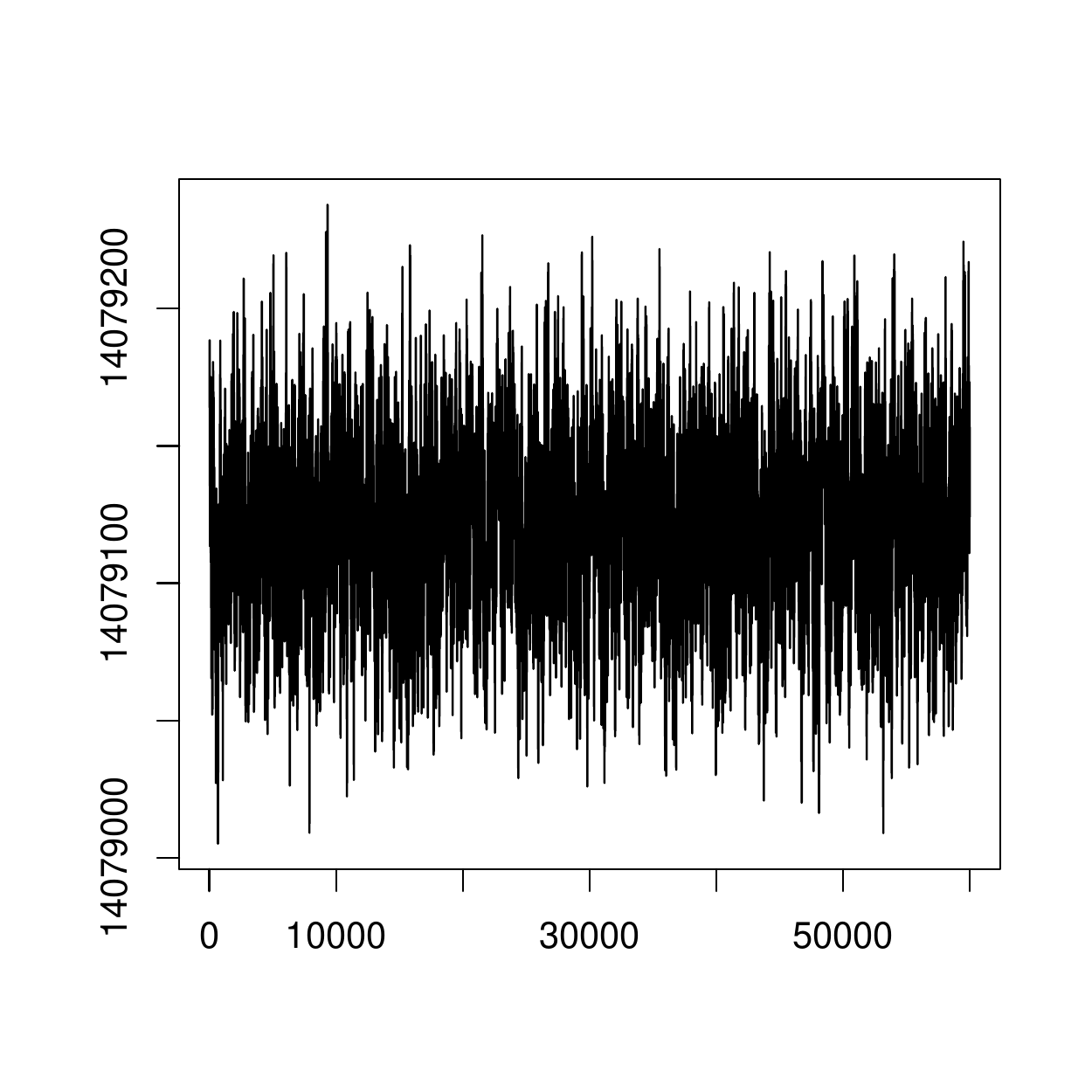}}
	\subfloat[][$\rho_{35}$]{\includegraphics[trim={.46in .65in .2in .78in},clip, totalheight=0.175\textheight]{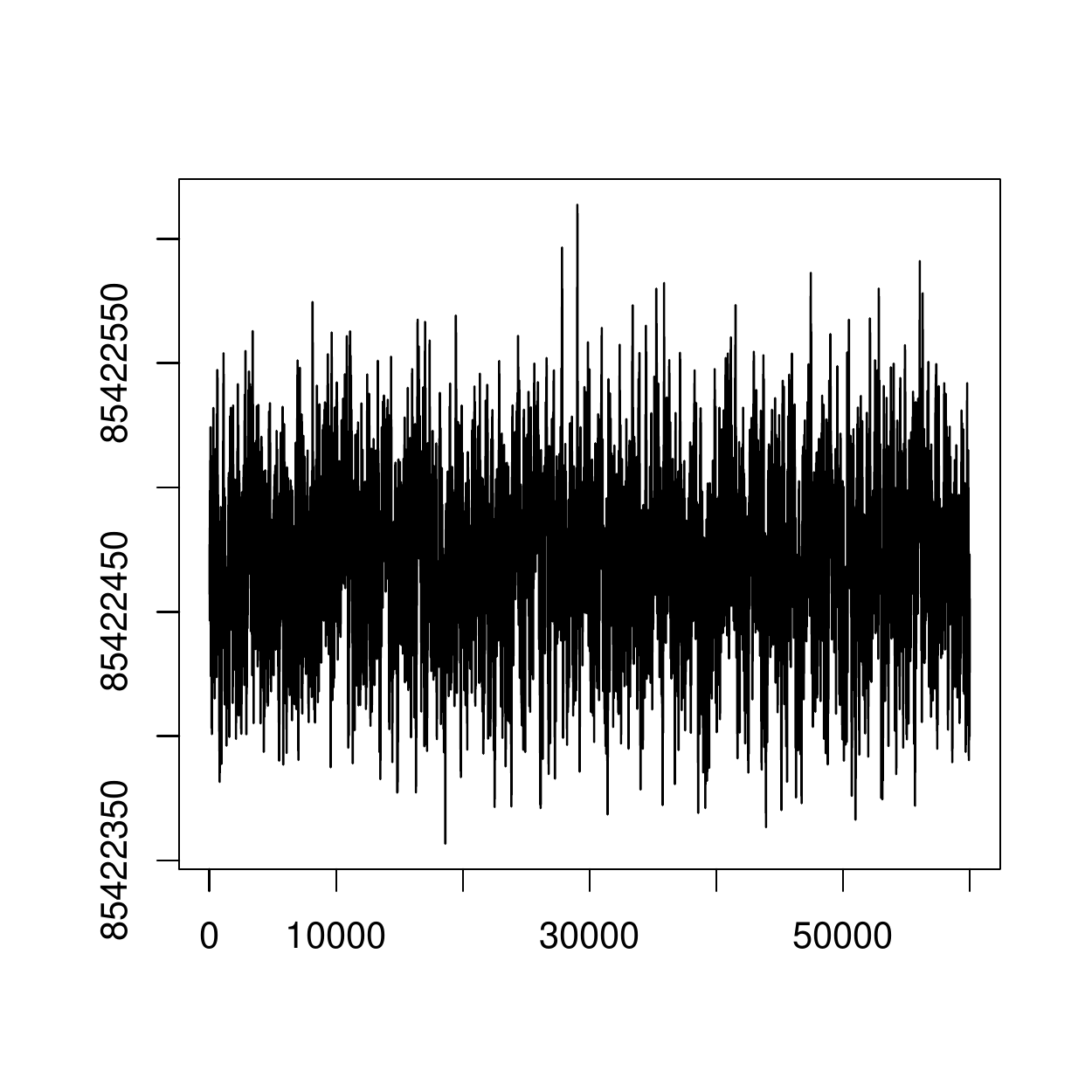}}	
	\caption{Traceplots of TMCMC samples.}
	\label{fig:traceplot}
\end{figure}

It is important to note that all the variables are updated through a single $\epsilon$ at each iteration and \ctp{Somak14} discussed its advantages especially in high dimensions.
\ctp{deyos16} discussed optimal scaling properties of the additive $TMCMC$ algorithm described in Algorithm \ref{algo:additive_TMCMC} where $g(\cdot)$ is a normal distribution with the negative part truncated, that is, $g(\epsilon)\equiv N(0,1)I_{\epsilon>0}$ 
for optimal acceptance rate of the algorithm and also prescribed a methodology to obtain approximate optimal scaling in practice. Following their prescription we have chosen the scaling 
parameters mentioned in Step 1 of Algorithm \ref{algo:additive_TMCMC} and generated samples from the distribution of our interest. As many as $1.5\times10^8$ samples were generated 
out of which the first $3\times10^7$ samples were discarded as burn-in. A TMCMC sample is stored at every $100^{th}$ iteration. Traceplots of some selected parameters shown in 
Figure \ref{fig:traceplot} provide evidence towards excellent mixing.

\subsection{Model validation}
\label{subsec:goodness}
\begin{figure}
	\centering
	\includegraphics[trim={.46in .64in .42in .78in},clip, totalheight=0.225\textheight]{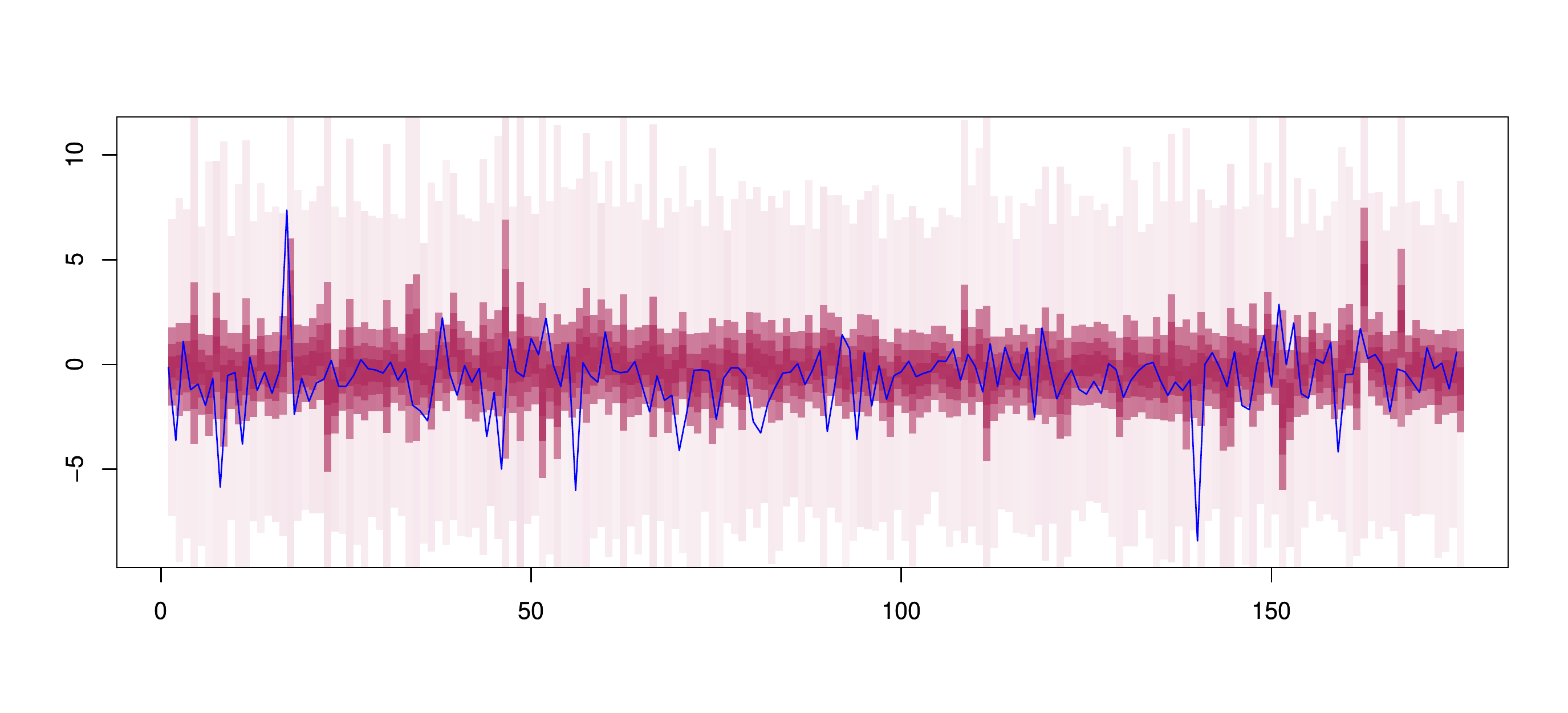}\\
	\includegraphics[trim={.46in .64in .42in .78in},clip, totalheight=0.225\textheight]{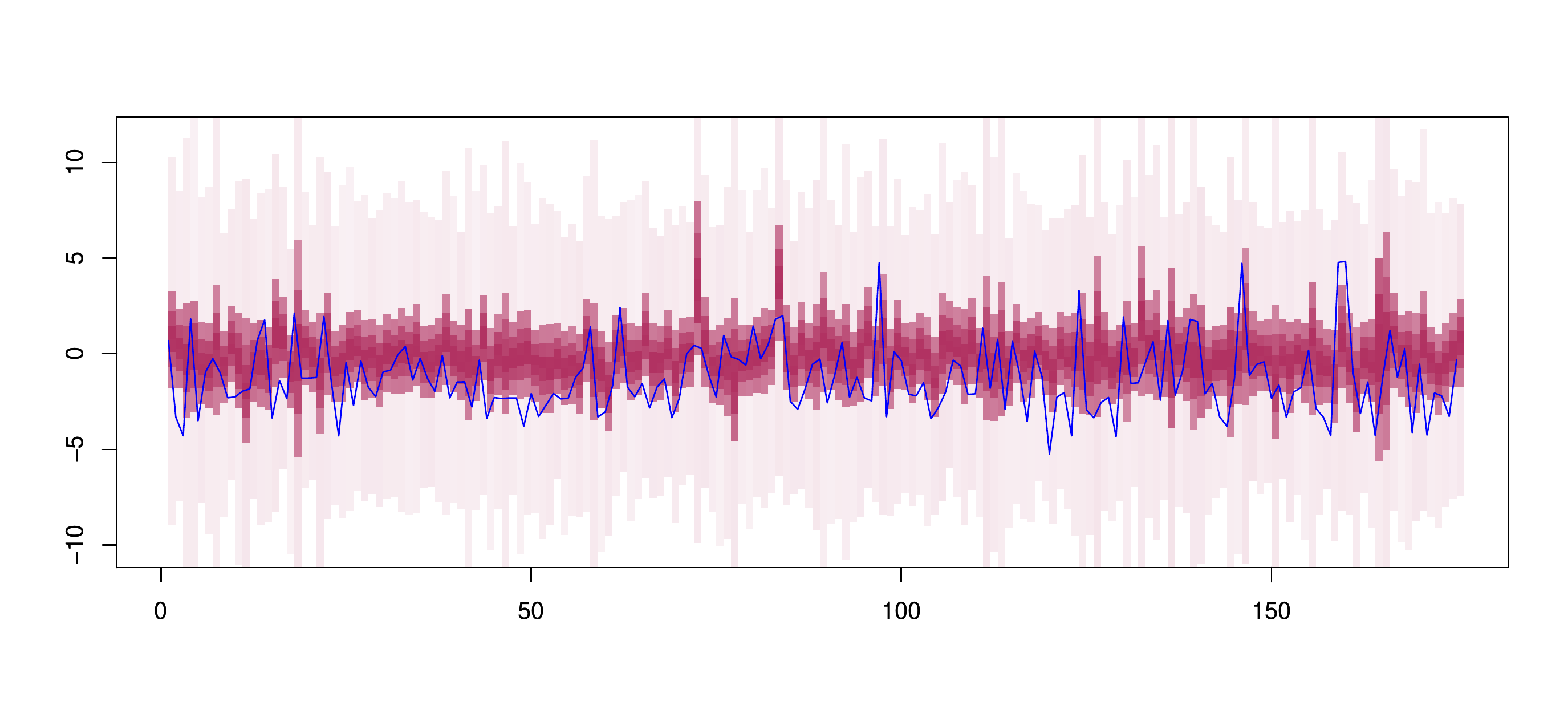}
	\includegraphics[trim={.46in .64in .42in .78in},clip, totalheight=0.225\textheight]{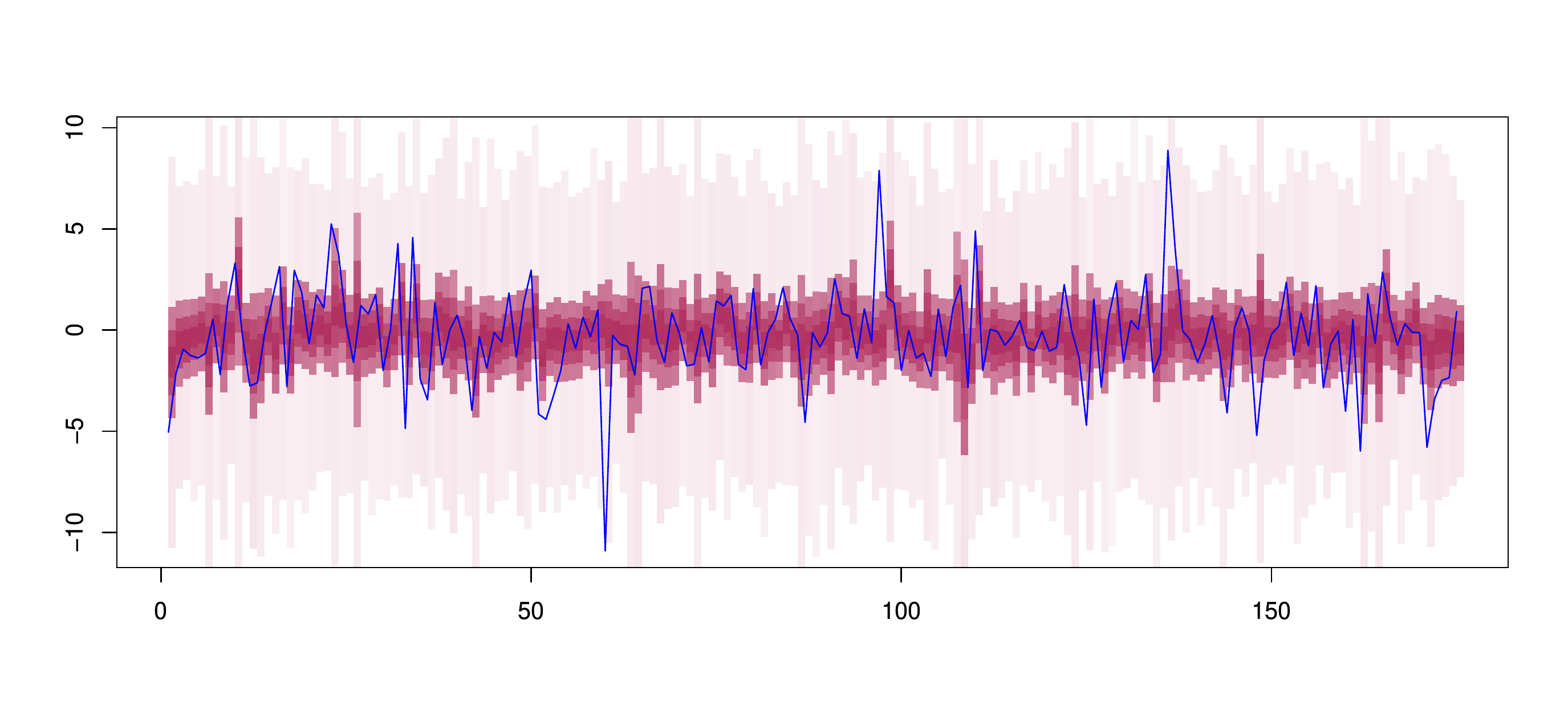} \\
	\includegraphics[trim={.46in .64in .4in .78in},clip, totalheight=0.225\textheight]{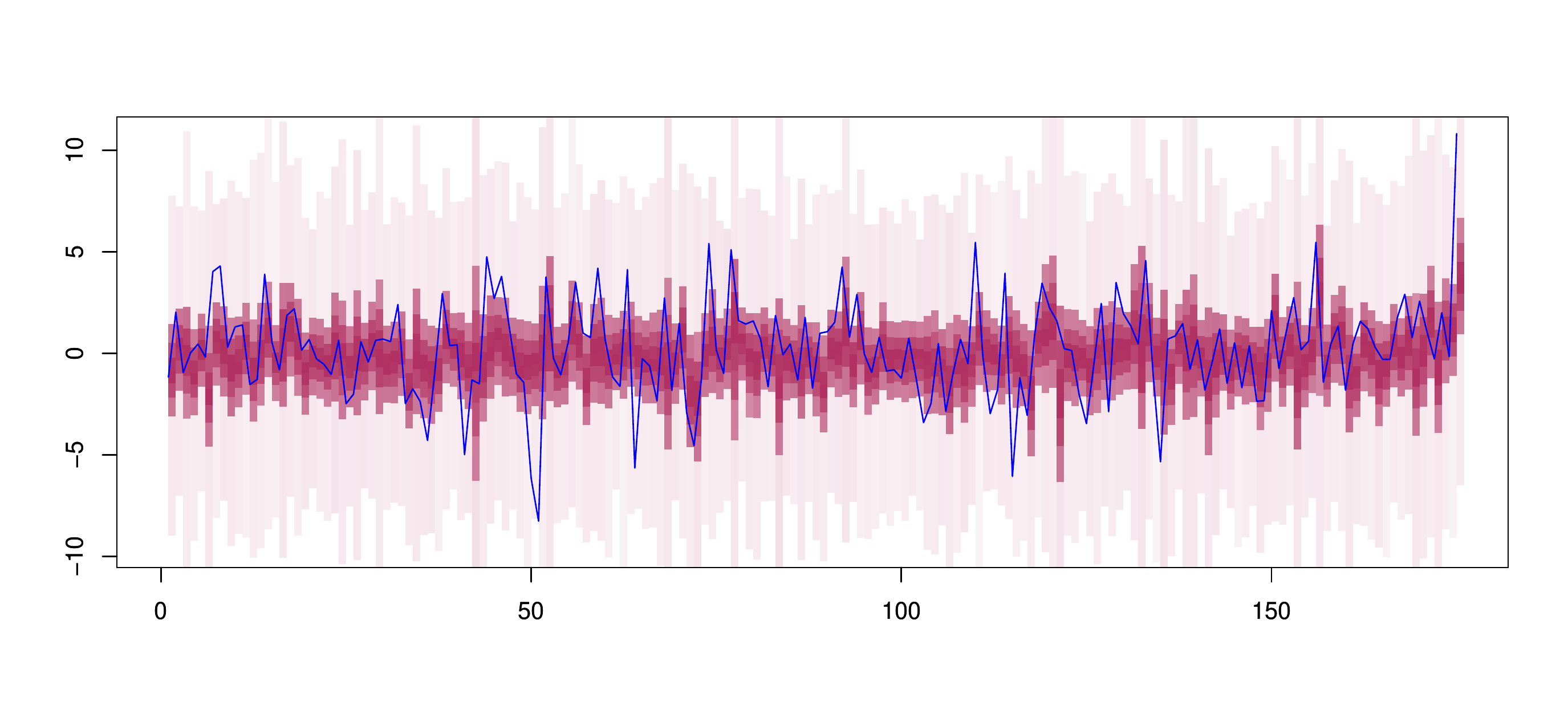} 	
	\caption{Leave-one-out cross validation.}
	\label{fig:cvplot}
\end{figure}

To validate our proposed model we conduct leave-one-out cross-validation for each of the $n$ data points by successively excluding the $i^{th}$ (for $i=1,\ldots,n$) data point and 
computing the corresponding posterior predictive distribution based on the remaining $n-1$ data points. The posterior predictive distributions are also approximated by 
drawing samples by the TMCMC algorithm. In Figure \ref{fig:cvplot} we show the results associated with four data points. The dark coloured region represents the $75\%$ 
credible region of the posterior predictive distribution of the $Z_{ij}$'s. We see that the true data point lies well within the credible region in most of the cases, even though the 75\%
credible regions are much narrow compared to the traditional 95\% regions that are usually advocated in general Bayesian analysis as a rule of thumb. On the basis 
of our leave-one-out posterior predictive results, we conclude that our postulated model very ably explains the variability of the data.

\section{Discussion on multiple testing methodology}
\subsection{Choice of $G_1,\ldots,G_m$} 
\label{sec:group_choice}
In Section \ref{subsec:motivation}, with proper biological motivation we have discussed that miRNAs generated from nearby locations have high chance to be coexpressed together. 
Thus, it is natural to form groups with miRNAs with nearby locations. Now as regards the statistical dependence among the miRNAs, we have considered the 
Mat\'ern covariance function $c^{(i)}(\cdot,\cdot)$ within strand $i$. Notably, 
$$c^{(i)}(x_j,x_l)\propto\frac{1}{\abs{x_j-x_l}}$$
for the genomic locations $x_j$ and $x_l$ where $\abs{x_j-x_l}$ is the distance between them. Thus forming groups of miRNAs on the basis of genomic distance is equivalent to forming groups based on prior correlation.

In practice, we first estimate the prior correlation matrix of $\bpsi$ 
from the model that we propose using the Monte Carlo method. Let $\bR$ be the correlation matrix 
with the $(i,j)$ element $r_{ij}$. We then consider the correlations between the $i^{th}$ and $j^{th}$ parameters, with $i < j$, and obtain the $95^{th}$ percentile, $r$, of these quantities. 
Then, in $G_i$ we include only those indices $j (\neq i)$ such that $r_{ij} \geq r$. Thus, the $i^{th}$ group contains indices of the parameters that are highly correlated with 
the $i^{th}$ parameter. If there exists no index $j$ such that $r_{ij} \geq r$, then we set $G_i = \{i\}$. To reduce complexity we restrict the groups to contain at most 
5 indices associated with at most 5 largest values of $r_{ij}$ exceeding $r$. This strategy has produced excellent results in the simulation studies by \cite{chandra2017}. It was observed that limiting the group size to 5 increases the power without much affecting the method complexity.
Therefore we employ this strategy in this application as well.

\subsection{Error measures in multiple testing}
\label{subsubsec:error_rates}
The \textit{False Discovery Rate} $(FDR)$ is defined as
\begin{equation}
FDR=E_{\bZ} \left[ \sum_{\dec\in\mathbb{D}}  
\frac{\sum_{i=1}^{m}d_i(1-r_i)}{\left\lbrace \sum_{i=1}^{m}d_i\right\rbrace \vee 1}\delta(\dec|\bZ) \right],
\label{eq:fdr}
\end{equation}
where $\delta(\dec|\bZ)$ is the probability of choosing the decision configuration $\dec$ 
according to the associated multiple testing procedure given data $\bZ$. In case of non-randomized decision rules, 
$\delta(\dec|\bZ)=1$ for the decision configuration which is chosen to be the final decision rule. Notably, given a particular data set, the non-marginal procedure also gives a binary vector as the optimal decision configuration, not any randomized decision rule.

Under the prior distribution of $\bpsi$, the \textit{posterior} $FDR$ is defined as :
\begin{align}
posterior~FDR
=&\pexp \left[ \sum_{\dec\in\mathbb{D}}  
\frac{\sum_{i=1}^{m}d_i(1-r_i)}{\left\{\sum_{i=1}^md_i\right\} \vee 1}\delta(\dec|\bZ) \right] =
\sum_{\dec\in\mathbb{D}}  \frac{\sum_{i=1}^{m}d_i(1-v_i)}{\left\{\sum_{i=1}^md_i\right\}\vee1}\delta(\dec|\bZ).
\label{eq:Psfdr}
\end{align}
These can be regarded as measures of Type-I error in multiple testing. Similarly \textit{False Non-Discovery Rate} $(FNR)$ stands as a measure of Type-II error. It is defined as:
\begin{equation}
FNR=E_{\bZ} \left[ \sum_{\dec\in\mathbb{D}}  
\frac{\sum_{i=1}^{m}(1-d_i)r_i}{\left\lbrace \sum_{i=1}^{m}(1-d_i)\right\rbrace \vee 1}\delta(\dec|\bZ) \right],
\label{eq:fnr}
\end{equation}
and the \textit{posterior} $FNR$ is 
\begin{align}
posterior~FNR
=&\pexp \left[ \sum_{\dec\in\mathbb{D}}  
\frac{\sum_{i=1}^{m}(1-d_i)r_i}{\left\{\sum_{i=1}^m(1-d_i)\right\} \vee 1}\delta(\dec|\bZ) \right] =
\sum_{\dec\in\mathbb{D}}  \frac{\sum_{i=1}^{m}(1-d_i)v_i}{\left\{\sum_{i=1}^m(1-d_i)\right\}\vee1}\delta(\dec|\bZ).
\label{eq:Psfnr}
\end{align}

As error criteria in Bayesian multiple testing paradigm, \ctp{chandra2017} advocated controlling posterior versions of the errors. In keeping with the Bayesian philosophy, 
it makes sense to control error measures conditional on the data, avoiding expectation with respect to the (marginal) distribution of the data. Not only does this support the 
Bayesian philosophy, it also drastically simplifies the computation of such error measures in complex practical problems. \ctp{fan2017} proposed a methodology for estimating $FDR$ 
under unknown arbitrary dependence. However, in their model there is no dependence structure between the hypotheses {\it a priori} and they also assumed sparsity. Such assumptions 
are not valid with respect to our realistic model and in general realistic composite hypothesis testing problems like ours. On the other hand, the posterior versions are readily 
available from TMCMC samples drawn from the posterior distribution. Therefore, we control \textit{posterior} $FDR$ and also estimate posterior $FNR$ for the obtained decision rule.

\section{Approximation of the $p$-values for the composite hypothesis testing problem}
\label{subsec:lrstat}
For BH adjustment of multiplicity correction, $p$-values are required corresponding to each hypothesis.  Note that
\begin{equation*}
Z_{i1},\ldots, Z_{in}\overset{iid}{\sim} \mathcal{N}(\psi_i,\sigma_i^2),~ i=1,\ldots, m.
\end{equation*}

The likelihood is given by:
\begin{equation*}
L_i(\psi_i,\sigma_i^2) = \prod_{j=1}^n f(Z_{ij};\psi_i,\sigma_i^2)
\end{equation*}
where $f(\cdot;\psi_i,\sigma_i^2)$ is the density of a standard normal distribution with mean $\psi_i$ and variance $\sigma_i^2$.

The LR-test statistics for the $i^{th}$ test is		
\begin{align*}
\zeta_i=&\frac{\underset{ \psi_i\in[-1,1],\sigma_i^2\in\mathbb{R}_+} {\sup} L_i(\psi_i,\sigma_i^2)} {\underset{ \psi_i\in\mathbb{R},\sigma_i^2\in\mathbb{R}_+} {\sup} L_i(\psi_i,\sigma_i^2)}
\end{align*}

The corresponding $p$-value is given by:
\begin{equation*}
p_i=P_{H_{0i}} \left(\zeta_i<\zeta^{(obs)}_i \right),
\end{equation*}	
where $\zeta^{(obs)}_i$ is the observed value associated with the random variable $\zeta_i$.
The null distributions of the $\zeta_i$ is not available in closed form and therefore we estimated $p_i$ using the bootstrap method and subsequently identified miRNAs with 
statistically significant differential expressions by the BH procedure. 
\bibliographystyle{natbib}
\bibliography{irmcmc}


\end{document}